# $MoS_2$ thin-film transistors spin-states, of conduction electrons and vacancies, distinguished by operando electron spin resonance

Naho Tsunetomo[1], Shohei Iguchi[1], Małgorzata Wierzbowska[2], Akiko Ueda[3,4], Won Yousang[1], Heo Sinae[5,6], Jeong Yesul[5], Yutaka Wakayama[5,6], Kazuhiro Marumoto[1,6]

[1]*Division of Materials Science, University of Tsukuba, Tsukuba, Ibaraki 305-8573, Japan*

[2]*Institute of High Pressure Physics, ul. Sokołowska 29/37, 01-142 Warsaw, Poland*

[3] *Spintronics Research Center, National Institute of Advanced Industrial Science and Technology (AIST) Tsukuba, 305-8568, Japan*

[4]*Department of Electrical Engineering, Columbia University, New York, NY 10027, USA*

[5]*International Center for Materials Nanoarchitectonics (MANA), National Institute for Materials Science (NIMS), Ibaraki 305-0044, Japan*

[6]*Department of Chemistry and Biochemistry, Faculty of Engineering, Kyushu University, 1-1 Namiki, Tsukuba 305-0044, Japan*

[6]*Tsukuba Research Center for Energy Materials Science (TREMS), University of Tsukuba, Ibaraki 305-8570, Japan*

*Correspondence should be addressed to marumoto@ims.tsukuba.ac.jp

**Transition metal dichalcogenide $MoS_2$ is a two-dimensional material, attracting much attention for next-generation applications thanks to rich functionalities stemmed from the crystal structure. Many experimental and theoretical works have focused on the spin-orbit interaction which couples the valley and spin degree of freedom so that the spin-states can be electrically controllable. However, the spin-states of charge carriers and vacancies have not been yet elucidated directly from a microscopic viewpoint. We report the spin-states in thin-film electric double-layer transistors using operando electron spin resonance (ESR) spectroscopy. We have observed clearly different ESR signals of the conduction electrons and vacancies, and distinguished the corresponding spin-states from the signals and theoretical calculations, evaluating the gate-voltage dependence and spin-susceptibility and *g*-factor temperature dependence. This analysis gives deep insight into the $MoS_2$ magnetism and clearly indicates the lower charge mobilities compared to graphene, which would be useful for improvements of the device characteristics and new applications.**

Since graphene was discovered in 2004, two-dimensional (2D) layered materials have attracted much attention. Among them, transition metal dichalcogenides (TMDs) denoted as $MX_2$ have been extensively studied. TMDs, with their representative material $MoS_2$, possess desired electrical and optical properties such as high mobility, circular-polarized light emission and flexibility, thus, deserving wide investigation also in the field of valleytronics[1,2,3,4,5]. The bandgap in $MoS_2$ single layer is direct, while it becomes indirect with increasing number of layers[6]. Since the nonzero bandgap exists in both the single layer and bulk, $MoS_2$ is expected to be applicable in the thin-film devices exceeding those with graphene[3,7,8,9,10]. High electric-field effect can be applied using ionic liquid, and the electric field-induced superconductivity in $MoS_2$ has been observed at low temperatures[11]. It has been reported, that the aforementioned properties are caused by the conduction- and valence-bands splitting by the spin-orbit interaction (SOI)[12,13,14], weak electronic localization and spin-orbit scattering (SOS)[15], discussed from the theoretical and experimental viewpoints. The spin-states of atomic vacancies have been studied using the density functional theory (DFT)[16,17,18]. Although the spin-states investigation is an important issue for the fundamental understanding of $MoS_2$ and applications of its magnetic properties, this matter has not been yet fully elucidated experimentally[19].

ESR spectroscopy is useful for the spin-states study of the organic electronic devices, such as transistors, solar cells and light-emitting diodes[20,21,22]. From the ESR study of graphene transistors, we have found a correlation between vacancies and their conducting mechanisms[23]. That is, spins exist in the charge-neutral graphene due to vacancies, however, when positive or negative gate-voltage is applied, spins vanish and then the spin-scattering of charge carriers decreases. Thus, the electrically-induced ambipolar spin-vanishment is believed to contribute to the high carrier-mobility in graphene[23]. Previous reports for $MoS_2$ thin films include the ESR and DFT studies of the atomic vacancies[16,17,18,19]. For $MoS_2$ transistors, the theoretical studies, fabrication methods and device-performance characterizations have been conducted[3,4,5,7,9,10,11], however, operando ESR study has not been yet performed. Elucidation of the spin-states of charge carriers and vacancies is important for understanding the physical properties and functioning of $MoS_2$ transistors.

Here we report $MoS_2$ thin-film transistors spin-states using the operando ESR. We have successfully observed several ESR signals, exhibiting various kinds of the electric-field response, and identified their origins as derived from the conduction electrons or atomic vacancies. The ESR signal-changes under the applied gate-voltage are compared with the *g*-factor variation calculated with the gauge-including projector-augmented wave (GIPAW) method[24]. The temperature-dependence of measured $g$-factors reflects the SOI and SOS, and is compared to the theoretical results using the Mori-Kawasaki formula[25]. Therefore, these studies give an important information on the microscopic properties of $MoS_2$ in the transistor structures.

In this work, we used a device configuration called the side-gate structure, schematically presented in Figure 1a with its cross section in Figure 1b. The $MoS_2$ thin film was formed on the sapphire ($Al_2O_3$) substrate with a multi-step chemical vapor deposition (CVD) method[26,27]. The gate, source and drain electrodes of Ni/Au (1/49 nm) were fabricated on $Al_2O_3$ or $MoS_2$ with a vacuum deposition method. The insulating layer of the transistor structure was formed by the ion-gel electric double-layers (EDLs), which enable to obtain a higher charge-density at the low gate-voltages than that of the conventional solid insulating layers[28]. The ion-gel was formed on the polyethylene terephthalate (PET) substrate by a drop casting and thermally annealed under the vacuum conditions. Then, the whole system was annealed and laminated in the $N_2$-filled glove box, and wired using the Ag paste. The fabricated transistor was inserted into the ESR sample tube and sealed under He atmosphere, as presented in Figure 1c. Details of the device fabrication method are described in Method section.

The spin-states in the device were observed with the operando ESR spectroscopy, using the liquid-He cryostat which enables to perform measurements from 4 K to room temperature. The gate-voltage ($V_G$) and drain-voltage ($V_D$) were controlled by an analyzer or a source meter, such that the operando ESR signals were measured simultaneously with the drain-current ($I_D$) and gate-current ($I_G$). Figure 1d shows the typical transfer characteristics of the fabricated device. The $I_D$ increased with the $V_G$ being positive and the n-type behavior was confirmed. The field-effect mobility $\mu$ = 6.6 cm$^2$ V$^{-1}$ s$^{-1}$ and on/off ratio of 436 were obtained as the transistor parameters. Method section contains more details of the characterization.

The spin-states were clearly reflected in the ESR signals and varied due to the electron accumulation under applied $V_G$, in “on” and “off” states. In addition, it was considered that the signals derived from the conduction electrons and atomic vacancies might exhibit different temperature dependencies. The former signals were expected to show the Pauli paramagnetism and the latter to obey the Curie law. The measured ESR spectra are presented in Figure 2, where three kinds of signals: A, B and C might be defined from the lowest applied magnetic field. The signals with the inverse phase at around 306, 314 and 323 mT are due to a standard $Mn^{2+}$ marker sample. We have confirmed that the signals at around 319 mT and 329 mT are attributed to the background signal of the ESR cavity resonator and the sapphire substrate, respectively.

Figure 2 shows a change of the signal intensity with $V_G$. To analyze this effect in detail, we compared the ESR spectra with $V_G$ switched on and off, and obtained the different signals reported in Figure 3. Figures 3a,c,e present the ESR spectrum of the Signal A, B and C, respectively. The red and blue lines show the spectra at $V_G$ = 2 V and 0 V, respectively. Figures 3b,d,f show the differences in the Signals A, B and C obtained at $V_G$ = 2 V with respect to the corresponding signals at $V_G$ = 0 V. Since the Signal A obtained at low temperatures overlapped with the background signal of the cavity resonator, we calculated each difference after correcting the corresponding baseline signal. The intensity of the Signal A increased with $V_G$, and the difference signal with $g$-factor of 2.0546 ± 0.004 was observed at 5 K. The Signal B also increased with $V_G$, and the difference signal with $g$ = 2.0026 ± 0.004 was obtained at 40 K. In contrast, the Signal C with $g = 1.9759 \pm 0.004$ was observed with $V_G = 0$ V and its intensity decreased with $V_G$. Therefore, it is demonstrated that these signals have different origins, because they show different $g$-factor and $V_G$ dependence.

Further insight is drawn from the temperature dependence of the $g$-factor and spin susceptibility ($\chi$), evaluated from the ESR spectra and presented in Figure 4. Specifically, the temperature dependence of the $\chi$ is presented in Figures 4a,c,e and of the $g$ values in Figures 4b,d,f for the Signals A, B and C, respectively. The spin susceptibility was evaluated for each ESR signal from the peak-to-peak ESR intensity $\Delta I_{pp}$ and the square of peak-to-peak ESR linewidth $\Delta H_{pp}$, that is $\Delta I_{pp}(\Delta H_{pp})^2$. Following, we derive the origins of the signals from the features of the $g$-factor and $\chi$.

We start with the Signal A. The $\chi$ shows almost no temperature dependence, as in Figure 4a, and this behavior is clearly different from that of the Curie law (dotted line in Figure 4a) which describes the isolated spins. Therefore, this signal is ascribed to be derived from the degenerated conduction electrons. Moreover, the $g$-factor generally determined by the SOI, decreased with temperature as in Figure 4b. The $g$-factor shift by the intrinsic SOI has been calculated using the DFT method and the value at the zero temperature in the absence of SOS has been reported to be $g^* \sim 2.21$ (refs. 29,30), which is larger than our experimental result ($g \sim 2.055$). This may imply that some spin-states are isolated at vacancies, while the majority of the spins may be located at the conduction band since the $g$-factors observed for the isolated spins are smaller than $g \sim 2$, as discussed below. Since it has been reported, that the spin-relaxation rate rises with increasing temperature[15], the $g$-factor shift might originate from the SOS. In the Supplementary Information (SI), we theoretically considered a simple model of the monolayer $MoS_2$, including the SOS and electron-phonon (e-ph) scattering, and assumed that the mechanism of the SOS was the Elliot-Yaffet type (EY) and D'yakonov-Perel' (DP) type, existing in the presence of the magnetic field[31]. The $g$-factor shift was derived from the calculation of the ESR response using the Mori-Kawasaki formula[25,32]. In Figure S1 in SI, we present a fit of the data in Figure 4b to the aforementioned model, using the parameters for the SOS and e-ph scattering rate given in Table S1 in SI. The result clearly shows, that the SOS assisted by the e-ph interaction lowers the $g$-factor. We observed that, the ESR signal lead to the different $g$-factor when the direction of the external magnetic field ($H$) was changed from the perpendicular to parallel with respect to the substrate plane. Since the orbital angular momentum is different for two orientations of $H$, the $g$-factor measured in the system is confirmed to be determined by the SOI.

For the Signal B, the $\chi$ exhibits the decreasing temperature dependence, as shown in Figure 4c. This behavior is characteristic for the Curie law for the isolated spins, thus, the signal is attributed to the vacancies in the $MoS_2$ thin films. As shown in Figure 4d, the $g$-factor also decreased with temperature. In addition, the $\Delta H_{pp}$ was obtained as 2.5 mT at 20 K and decreased monotonically to 0.8 mT at 60 K. Decreasing ESR linewidth with temperature indicates that the isolated spins are mobile, that is the motional narrowing of the ESR linewidth. In this case, the SOS effect is believed to increase with temperature. Thus, the SOS might contribute to the $g$-factor and lowers it with temperature, the same way as for the Signal A.

For the Signal C, akin to the Signal B, a Curie-like temperature dependence was observed, in which the $\chi$ increased with temperature, as shown in Figure 4e. Therefore, we ascribe this ESR signal to the isolated spins at vacancies in the $MoS_2$ thin films. No temperature dependence of the $g$-factor was observed, as shown in Figure 4f. In addition, the $\Delta H_{pp}$ did not vary much, changing from 1.7 to 2.0 mT in the range of 5 to 60 K. Therefore, the isolated spins are immobile (or localized), and the effect of the SOS might be small.

In order to investigate in detail the vacancy states in $MoS_2$, being in the origin of the Signal B and C, the DFT calculations were performed. Earlier works report several types of atomic vacancies, and among them, S, $S_2$, $MoS_3$ and $MoS_6$ vacancies have been discussed as the defects that easily occur in $MoS_2$ (refs. 16,17,33). Thus, we focus on these four types of vacancies. Figure S2 in SI displays examples of calculated spin-density distribution around the above vacancies in the $MoS_2$ monolayer, obtained with the DFT method, using the Quantum ESPRESSO package. To mimic the n-type operation of the ESR, in Figure S2 in SI, the spin-densities correspond to the negatively charge-doped systems. We introduce the defects and electron-doping to a monolayer (mL), bilayer (bL), trilayer (tL) and bulk material. For the bL, three configurations: A, B and C of the vacancy localization were considered, as shown in Figure S3 in SI. The electron-doping dependence of the magnetization was calculated for the mL, bL, tL and bulk $MoS_2$ without and with the vacancies, and it is presented in Figure S4 in SI. More detailed spin-density distribution maps are plotted in Figures S5-9 in SI. For the charge neutral defects, the local magnetization is non-vanishing only for the $MoS_6$ vacancy. Under the electron doping, the spin-density and magnetization decrease for the $MoS_6$ vacancy (see Figures S2d, S4c, and S9). In contrast, for the remaining vacancies: S, $S_2$ and $MoS_3$, as well as the absence of the defects, the spin-density and magnetization increase with the weak electron-doping (see Figures S2, S4, and S5-S8).

As mentioned above, for the charge-neutral film, the calculated spin-polarization vanishes for the cases of the S, $S_2$ and $MoS_3$ vacancies and no vacancy. This result agrees with the calculated electronic structures with non-vanishing bandgaps (Figures 5a-d). In contrast, for the $MoS_6$ vacancy, the bandgap disappeared for the spin down, indicating a semimetal state (as shown in Figures 5e,f), which corresponds to the existence of magnetization even for charge-neutral film.

We compare the experimental and calculated results. As can be seen from Figures 3e,f, the Signal C can be observed even without applied $V_G$, and the signal-intensity decreases with $V_G$. From the DFT results shown in Figure S4c in SI, only the $MoS_6$-vacancy state is magnetic without electron-doping, among the vacancies that have been thought to occur easily in $MoS_2$. Therefore, the Signal C is reasonably ascribed to be derived from the $MoS_6$ vacancy.

Contrary to the Signal C, the intensity of the Signal B increases with $V_G$ (see Figures 3c,d). Since the magnetization increases with $V_G$, three kinds of vacancies: S, $S_2$ and $MoS_3$, can be a plausible origin of the Signal B, examining the DFT results in Figure S4 in SI. To further identify the origin of this signal, the $g$-factors corresponding to these three vacancies were calculated, and the results are summarized in Figure 6. Also, the $g$-factor shifts, with respect to the free-electron's value (2.0023) in ppm units, and their dependence on the electron-doping are shown in Figure S10 in SI. As shown in Figure 6, the experimental $g$-factors of the Signal B, of 1.987-2.003, agree with the corresponding numbers obtained from the GIPAW method for the S-vacancy case. Moreover, the $g$-factors calculated for the $S_2$ and $MoS_3$ vacancies do not coincide with the experimental values. Therefore, the Signal B could be ascribed as associated with the S vacancy.

Our results for the $MoS_2$ thin films are compared with the previous ones for graphene. The mobility $\mu$ in graphene has been reported as about tens of thousands $cm^2\ V^{-1}\ S^{-1}$ (ref. 34), to be compared with values of several hundred to a thousand $cm^2\ V^{-1}\ S^{-1}$ for $MoS_2$ (ref. 35). Since carbon atoms of graphene have four valence electrons and three C-C bonds due to the C3 symmetry, the carbon vacancy necessarily has one unpaired electron leading to a nonzero spin-density[36]. In contrast, the unpaired electrons do not occur in $MoS_2$ even if atomic vacancies: S, $S_2$, $MoS_3$ are formed, as shown in this study. The reason might be as follows: Mo and S atoms have six and three Mo-S bonds, respectively, and six valence electrons. Thus, S atom has three extra valence electrons, not used in the single bonds. For the S, $S_2$ and $MoS_3$ vacancies, three, six and nine electrons remain unpaired, respectively. These unpaired electrons can be coupled with counter electrons by three extra valence electrons of S, which leads to the spin-unpolarized state. For the electron-doping, as described above, the spin-density never vanishes and is believed to make an effect on the spin-scattering of charge carriers, lowering the mobility. Contrary to $MoS_2$, in graphene, the ambipolar spin-vanishing under applied $V_G$ has been demonstrated, and discussed to cause an improvement of the charge mobility by a suppression of the spin-scattering of charge carriers[23]. Therefore, the investigations of the spin-states of the vacancies extend our knowledge on the mechanism of the high mobility emergence.

$MoS_2$ thin-film transistors have been fabricated for the studies presented here, and the n-type transistor operation has been confirmed. The operando ESR measurements under electron-doping have been performed with the variable $V_G$ and temperature. Three kinds of signals: Signal A, B and C, have been observed. The Signal A and B increase and the Signal C decreases with the applied $V_G$, respectively. For the Signal A, the spin-susceptibility is almost independent on temperature, indicating that the origin of this signal could be ascribed to the degenerated conduction electrons. For the Signal B and C, the spin-susceptibility shows the Curie-like temperature dependence, which means that the origins of these signals could be ascribed to the isolated spins. To further investigate the ESR-signals origins, the DFT and GIPAW calculations have been performed for four types of vacancies: S, $S_2$, $MoS_3$ and $MoS_6$, that are most likely to occur in $MoS_2$. As a result, only $MoS_6$ vacancy has a non-vanishing spin-density without the electron-doping, which enables us to identify the Signal C origin to be associated with the $MoS_6$ vacancy. In order to identify the origin of the Signal B, the observed

*g*-factor has been compared with that from the GIPAW calculations for S, $S_2$ and $MoS_3$ vacancies. As a result, only the *g*-factors calculated for the S vacancy agree with the ESR Signal B. Interestingly, the *g*-factors of the Signal A and B show the temperature dependencies, which we assign to an effect of the SOS in $MoS_2$ using the Mori-Kawasaki formula. In conclusion, the spin-states of the conduction electrons and vacancies (defects), being present in the $MoS_2$ thin film, have been clarified in this study by the operando ESR measurement during the device operation, to our knowledge for the first time. This result gives a deep insight into our understanding of the spin-states in $MoS_2$ from the microscopic point of view, and suggests a combined theoretical and experimental approach for other 2D TMDs.

## Methods

### Fabrication of $MoS_2$ thin film with multi-step CVD method

A sapphire ($Al_2O_3$) substrate (3 mm×20 mm) was cleaned by ultrasonic with acetone and 2-propanol, and then cleaned by UV ozone. A $MoO_3$ (99.9%, Sigma Aldrich) thin film of 3 nm was fabricated on the sapphire substrate with a vacuum deposition method under $1 \times 10^{-5}$ Pa. To crystalize the $MoO_3$ film, the film was annealed at 325°C for 120 min under $O_2$ atmosphere at 200 sccm (standard cubic centimeter per minute). Sulfur powder was placed on a quartz boat, which was placed side-by-side with the $MoO_3$ film in a quartz tube. The sulfur powder and the $MoO_3$ film were heated at 275°C and 550°C, respectively, under $N_2$ 50 sccm at atmospheric pressure for 60 min, which sulfurized the $MoO_3$ film to produce $MoS_2$. An annealing treatment was performed at 1000 °C for 30 min under Ar 200 sccm atmosphere to improve the crystallinity.

### Fabrication of transistors with $MoS_2$

Ni/Au (1/49 nm) layers were fabricated as electrodes on the $MoS_2$ film on the sapphire substrate by a vacuum vapor deposition method using an ULVAC VPC-260F system. An ion-gel solution was fabricated with ionic liquid 1-ethyl-3-methylimidazolium bis(trifluoromethylsulfonyl)imide ([EMIM][TFSI]) (36 wt%) (Ionic Liquids Technologies, Inc.), an ABA triblock polymer poly(styrene-*b*-methylmethacylate-*b*-styrene) (PS-PMMA-PS) (3 wt%) (Polymer Source, Inc.), and a solvent ethyl acetate (61 wt%) (Wako Pure Chemical Industries, Ltd.); the solution was stirred for more than 1 day. The ion-gel layer was formed on an ultrasonically cleaned polyethylene terephthalate (PET) substrate (3 mm×25 mm) by a drop-cast method, followed by a vacuum annealing treatment at 70 °C for more than 2 days. The ion-gel insulating layer has an electric double layer (EDL) and high ion conductivity[37,38,39]. The capacitance of the EDL is generally very large, causing a high charge-density state at low voltage and a high on/off ratio in transistor operation. After the vacuum annealing treatment, the ion-gel layer on the PET substrate and the $MoS_2$ thin film on sapphire substrate were transferred into a glove box ($O_2$ < 0.2 ppm, $H_2O$ < 0.5 ppm) and annealed at 70 °C for 2 h or more to remove adsorbed moisture. A wiring was performed using a silver paste in the glove box, and finally the ion-gel layer was laminated on the $MoS_2$ thin film. The fabricated transistor was placed into an ESR sample tube and sealed in it, where the inside of the sample tube was replaced with a He atmosphere.

**Measurements with device analyzer and ESR**

Transfer characteristics were measured with a device analyzer (KEYSIGHT B1500A). The $V_D$ was fixed at 1 V and the $V_G$ was swept from 0 to 1.5 V, where the data were measured every 30 mV per 60 s at each point. ESR measurements were performed with a X-band ESR spectrometer (JEOL RESONANCE JES-FA200) and a source meter (Keithley 2612A).

**Evaluation of spin concentration**

For the Signal B, the number of spins ($N_{spin}$) was evaluated by doubly integrating the difference ESR spectrum at 40 K and comparing it with a $Mn^{2+}$ marker. Considering the Curie low and the temperature correction at room temperature (297 K), the $N_{spin}$ was evaluated as $N_{spin}$ = $8.84\times10^{12}$. In the same way, the $N_{spin}$ of the Signal C was evaluated to be $N_{spin} = 49.15\times10^{12}$. The used active area of the transistor structure is 0.5 mm×12 mm = 6.0 $mm^2$ = $6.0\times10^{-2}$ $cm^2$. Since the $MoS_2$ thin film used in this experiment has five stacked layers, the spin concentration is calculated to be approximately 3% and 16% for Signal A and B, respectively, using the lattice constant of $MoS_2$ as 3.16Å.

**DFT calculations**

Density functional theory (DFT) calculations were performed using the Quantum ESPRESSO package[40], which is a plane-wave basis tool with the pseudopotentials describing the core electrons. For the $g$-tensor calculations, the norm-conserving gauge-including projector augmented wave (GIPAW) pseudopotentials were used[41,42], with the energy cut-off 60 Ry. The $g$-tensor was obtained with the QE-GIPAW post-processing code[24,43]. The uniform mesh of $k$-points was set to $12 \times 12 \times 1$ for the elementary cell, and $9 \times 9 \times 1$ and $6 \times 6 \times 1$ meshes for the $3 \times 3 \times 1$ and $4 \times 4 \times 1$ supercells, respectively. The vacuum separation between the periodic images of the 2D layers was set to 20 Å.

The band structures were interpolated using the maximally-localized Wannier functions (MLWF)[44] implemented in the wannier90 tool[45]. The spin-polarization maps were plotted with the XCrySDen tool[46].

**Acknowledgements**

DFT calculations were performed in the Cyfronet Computer Centre using the Prometheus computer. This work was partially supported by JSPS KAKENHI Grant Numbers JP19K21955 and JP19K05201, by JST PRESTO, by The Hitachi Global Foundation, by The MIKIYA Science and Technology Foundation, by The Futaba Foundation, and by JST ALCA Grant Number JPMJAL1603, Japan.

**Author contributions**

N.T. and K.M. conceived and designed this work. N.T., S.I., W.Y. and K.M. fabricated all the devices, performed the experiments and analyzed the data. M.W. performed the DFT calculations of $g$-factor and spin density distribution. A.U. performed the calculation of the effective $g$-factor at finite temperature. N.T., H.S., J.Y. and Y.W. fabricated all $MoS_2$ thin films. N.T., M.W., A.U. and K.M. wrote the manuscript. K.M. supervised this work. All authors discussed the results and reviewed the manuscript.

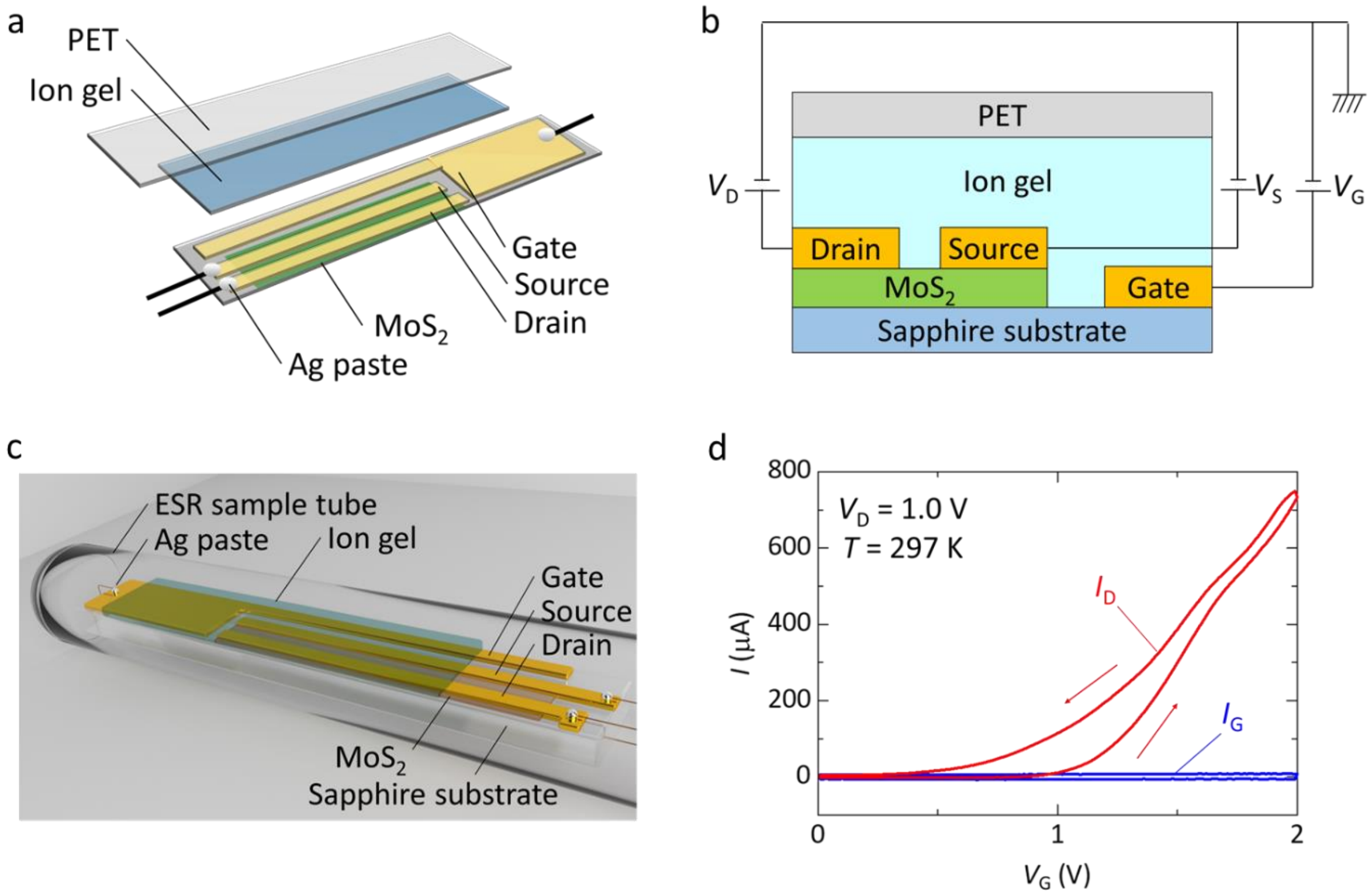

**Figure 1 | $MoS_2$ transistor structure and device characteristic.** **a**, Schematic diagram of the device structure of a $MoS_2$ transistor used for ESR measurements. **b**, Schematic cross section the device structure. **c**, Schematic diagram of the $MoS_2$ transistor in an ESR sample tube. **d**, Transfer characteristic of the $MoS_2$ transistor.

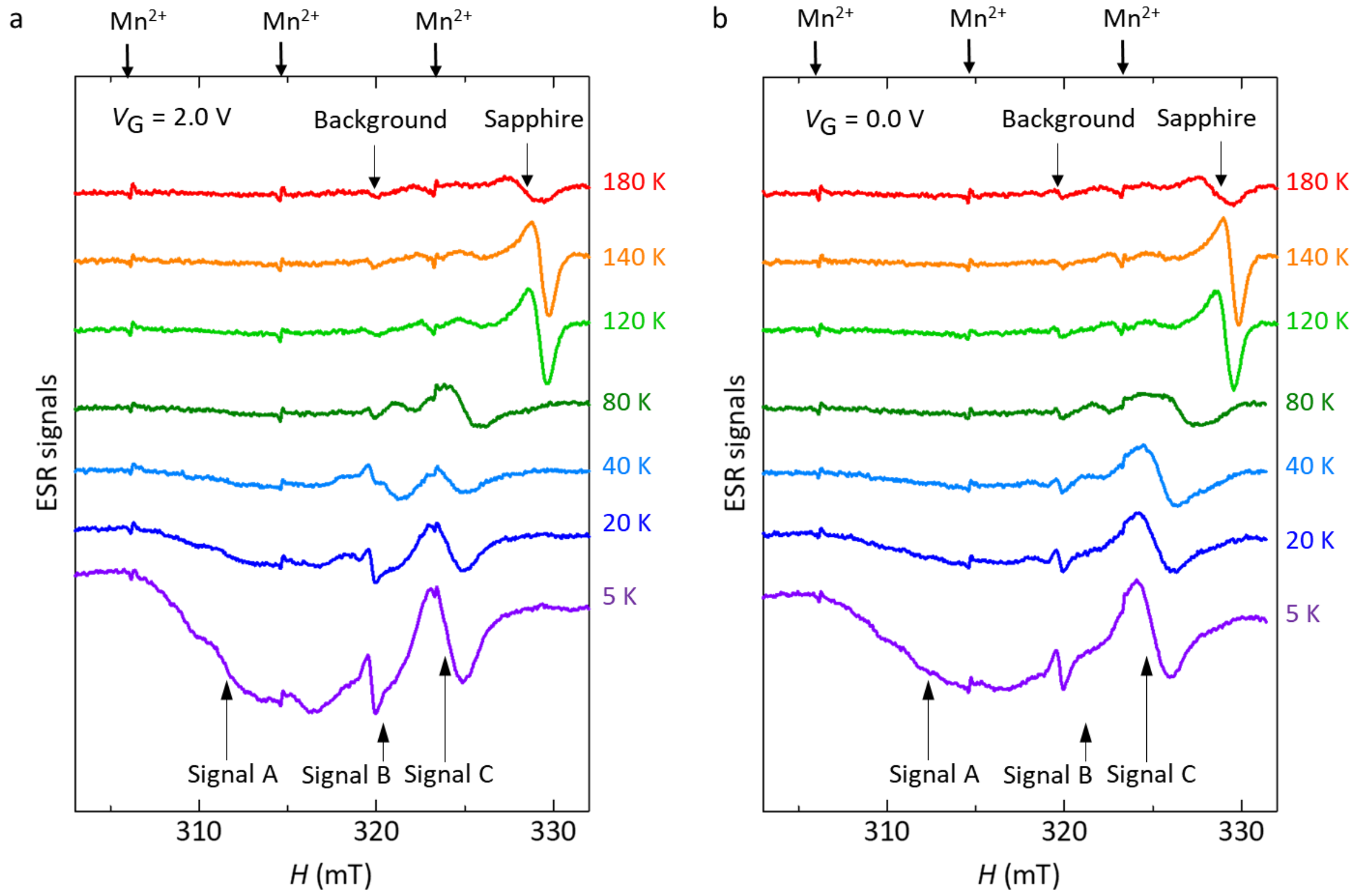

**Figure 2 | Temperature and gate-voltage dependence of ESR spectra of the $MoS_2$ transistor.** **a**, ESR spectra measured at a gate voltage ($V_G$) of $V_G = 2$ V and a drain voltage ($V_D$) of $V_D = 1$ V. **b**, ESR spectra measured at $V_G = 0$ V and $V_D = 1$ V. Distortion of the baseline of the spectra at low temperatures, typically below 40 K, is due to the background signal of the ESR cavity resonator.

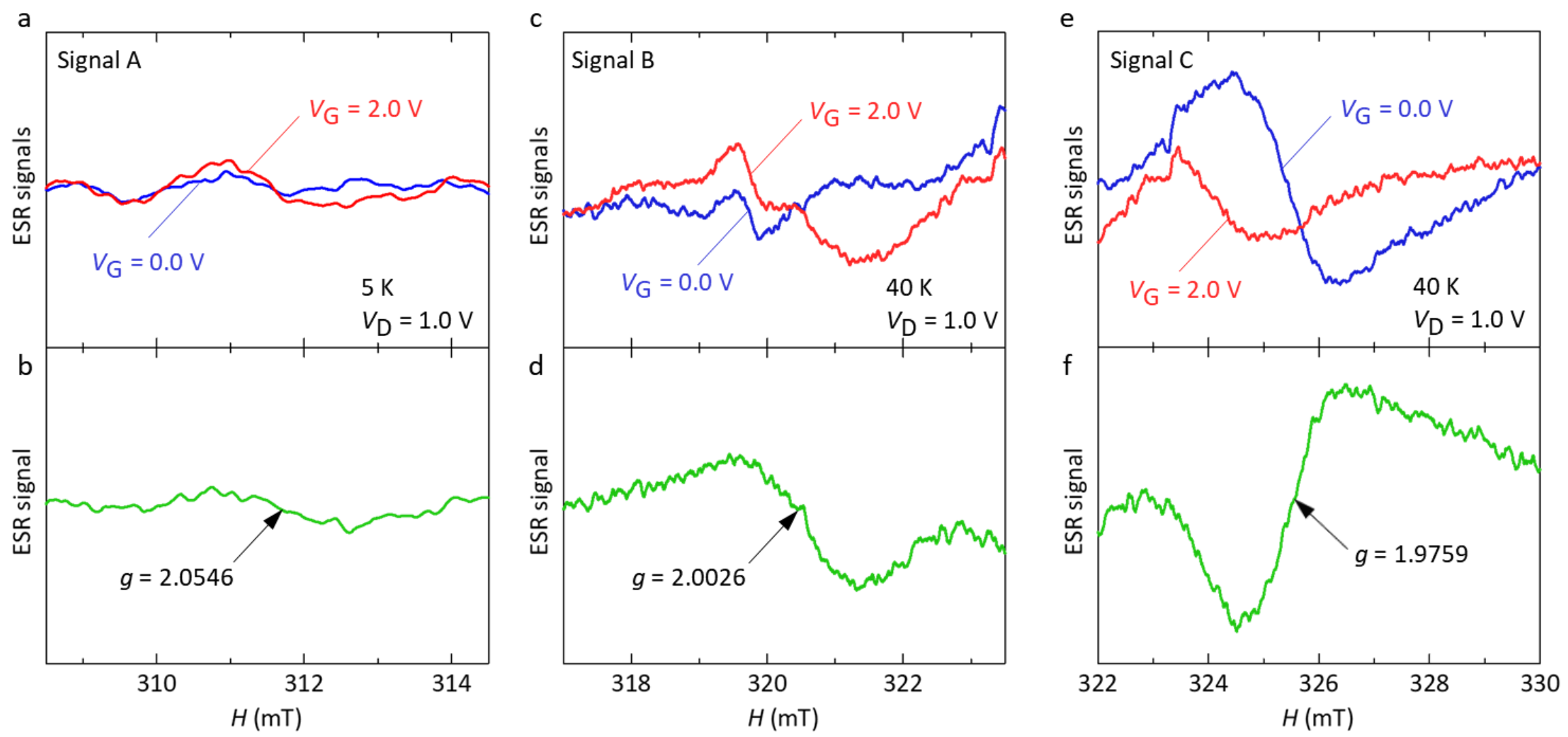

**Figure 3 | Difference ESR spectra by applying gate voltage.** The red lines show the spectra at $V_G = 2$ V and $V_D = 1$ V. The blue lines show the spectra at $V_G = 0$ V and $V_D = 1$ V. The green lines show the difference spectra obtained by subtracting the ESR spectra at $V_G = 0$ V from that at $V_G = 2$ V. **a**, Enlarged view of the ESR spectra of the Signal A at 5 K shown in Figure 2. Baseline correction has been performed on the background signal from the ESR cavity resonator. **b**, Difference ESR spectrum of the Signal A shown in Figure 3a. **c**,**e**, Enlarged views of the ESR spectra of the Signal B and C shown in Figure 2, respectively. **d**,**f**, Difference ESR spectrum of the Signal B and C shown in Figure 3c,e, respectively.

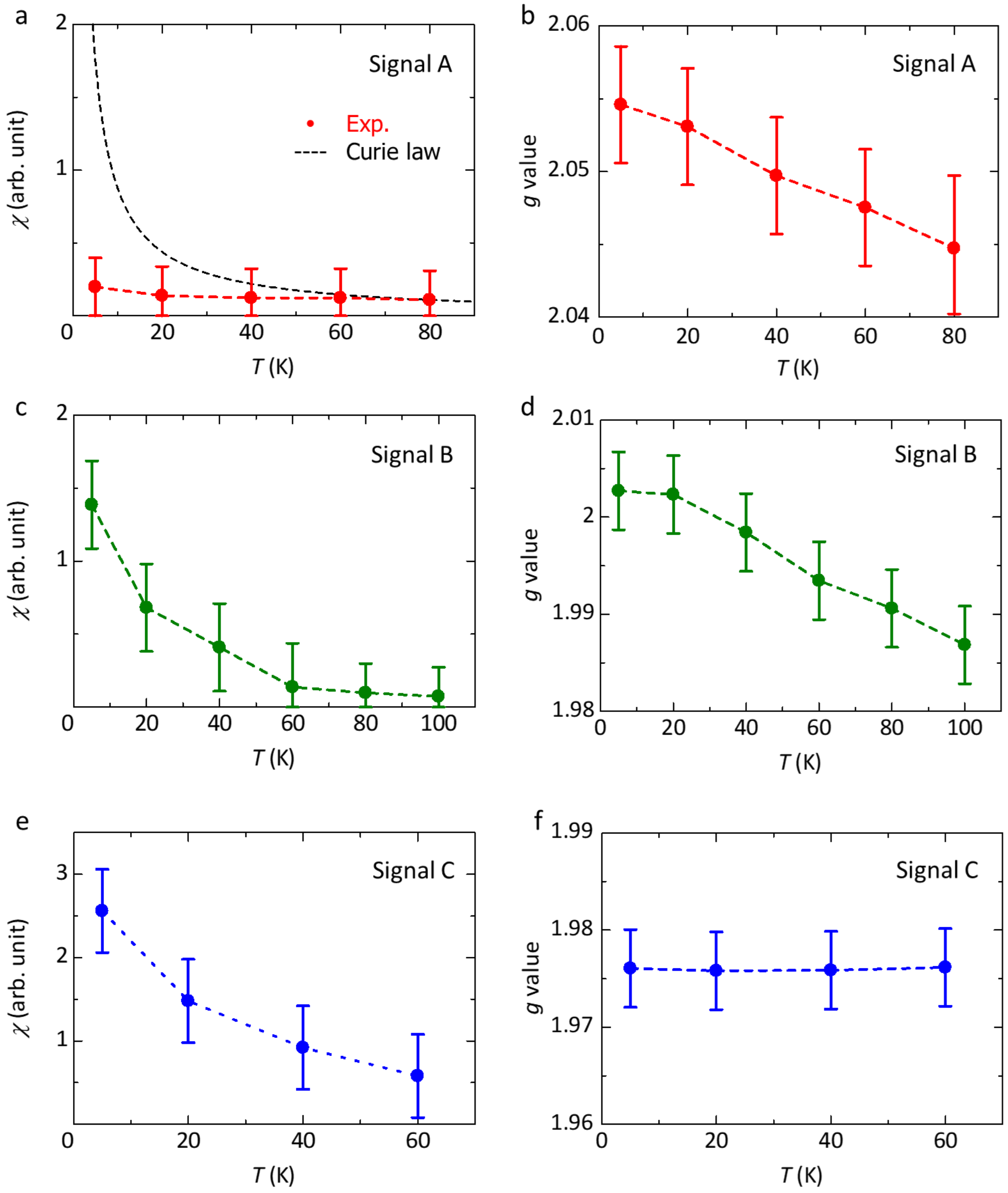

**Figure 4 | Temperature dependence of spin susceptibility and *g*-factor.** **a**,**c**,**e**, Temperature dependence of the spin susceptibility ($\chi$) of the Signal A, B and C, respectively. **b**,**d**,**f**, Temperature dependence of the *g*-factor of the Signal A, B and C, respectively.

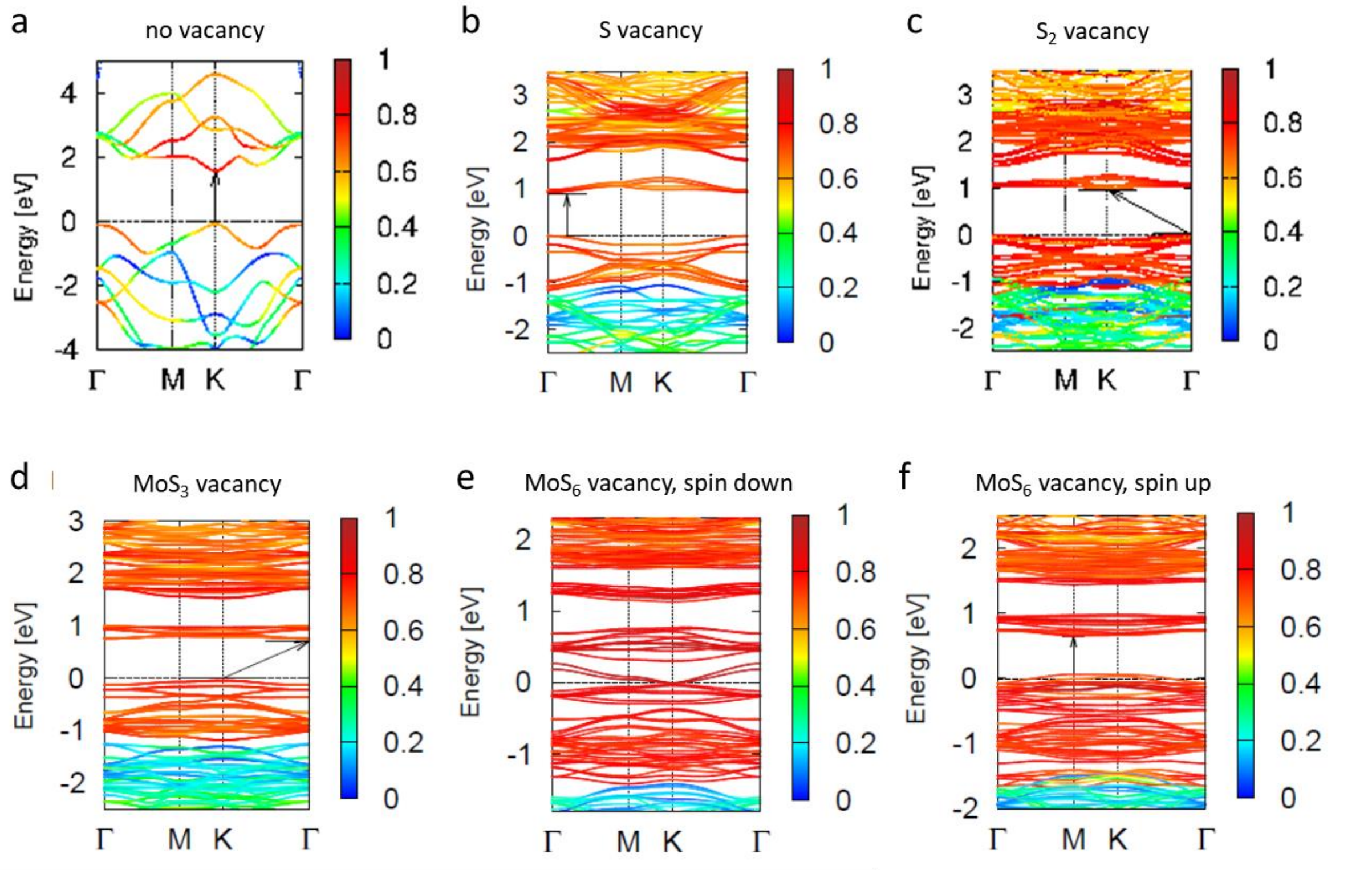

**Figure 5 | Calculated band structures of mL $MoS_2$ without and with various vacancies under no charge injection.** **a**, $MoS_2$ with no vacancy. **b**, $MoS_2$ with the S vacancy. **c**, $MoS_2$ with the $S_2$ vacancy. **d**, $MoS_2$ with the $MoS_3$ vacancy. **e**,**f**, $MoS_2$ with the $MoS_6$ vacancy for (e) spin-down and (f) spin-up case, respectively. When the electron doping is zero, no spin density or no magnetization is calculated for the cases of the S, $S_2$ and $MoS_3$ vacancies, as shown in Figure S4. These results are further confirmed by the existence of the band gap at the Fermi level with no spin density as shown in (b), (c) and (d). For the $MoS_6$ vacancy, the band gap is closed at the Fermi level depending on the spin direction as shown in (e) and (f), which means a semimetal state. The color legend represents the projections of the bands on the Wannier functions localized at the Mo atoms.

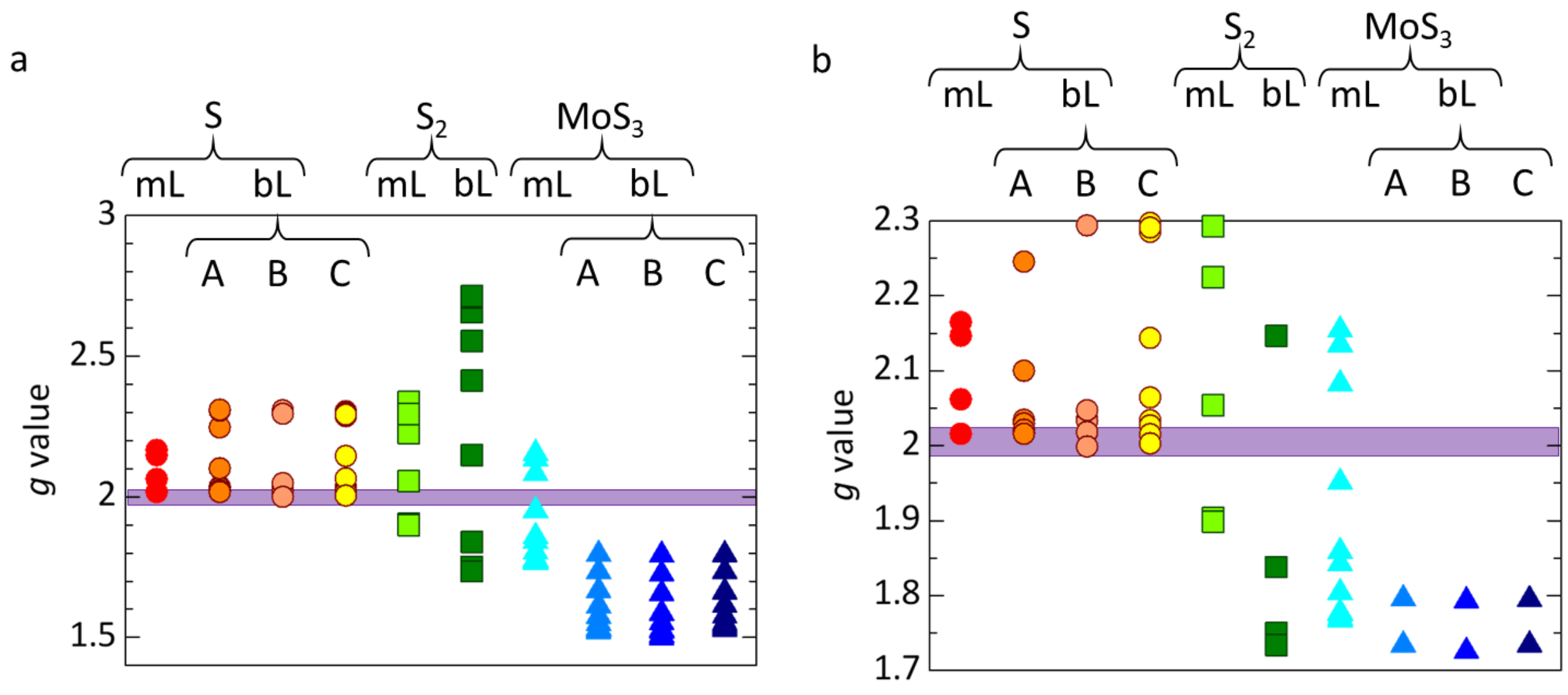

**Figure 6 | Calculated *g*-factor for various vacancies and charge doping.** The *g*-factor is calculated for the S, $S_2$ and $MoS_3$ vacancies in the $MoS_2$ supercell of the size 12 × 12 × 1 unit, with 1 to 12 injected electrons, are collectively displayed. Purple area indicates the range of the observed *g*-factors of the Signal B ($g$ = 1.987 to 2.003). For the case of bilayer (bL), type A, B and C are defined depending on the arrangement of vacancies (See Figure S3). These *g*-factors are calculated for the direction of external magnetic field perpendicular to the $MoS_2$ plane (*z*-axis) (see Figure S10). **a**, All calculated *g*-factors for the Signal B. **b**, Enlarged view near the observed *g*-factors of the Signal B ($g$ = 1.987 to 2.003).

## References

1. Song, I., Park, C. & Choi, H. C. Synthesis and properties of molybdenum disulphide: From bulk to atomic layers. *RSC Adv.* **5**, 7495–7514 (2015).
2. Xiao, D., Liu, G. Bin, Feng, W., Xu, X. & Yao, W. Coupled spin and valley physics in monolayers of $MoS_2$ and other group-VI dichalcogenides. *Phys. Rev. Lett.* **108**, 196802 (2012).
3. Kim, S. *et al.* High-mobility and low-power thin-film transistors based on multilayer $MoS_2$ crystals. *Nat. Commun.* **3**, 1011–1017 (2012).
4. Wang, Q. H., Kalantar-Zadeh, K., Kis, A., Coleman, J. N. & Strano, M. S. Electronics and optoelectronics of two-dimensional transition metal dichalcogenides. *Nat. Nanotechnol.* **7**, 699–712 (2012).
5. Pu, J., Li, L. J. & Takenobu, T. Flexible and stretchable thin-film transistors based on molybdenum disulphide. *Phys. Chem. Chem. Phys.* **16**, 14996–15006 (2014).
6. Splendiani, A. *et al.* Emerging photoluminescence in monolayer $MoS_2$. *Nano Lett.* **10**, 1271–1275 (2010).
7. Radisavljevic, B., Radenovic, A., Brivio, J., Giacometti, V. & Kis, A. Single-layer $MoS_2$ transistors. *Nat. Nanotechnol.* **6**, 147–150 (2011).
8. Ellis, J. K., Lucero, M. J. & Scuseria, G. E. The indirect to direct band gap transition in multilayered MoS2 as predicted by screened hybrid density functional theory. *Appl. Phys. Lett.* **99**, (2011).
9. Lembke, D. & Kis, A. Breakdown of high-performance monolayer $MoS_2$ transistors. *ACS Nano* **6**, 10070–10075 (2012).
10. Yin, Z. *et al.* Single-layer $MoS_2$ phototransistors. *ACS Nano* **6**, 74–80 (2012).
11. Ye, J. T. *et al.* Superconducting dome in a gate-tuned band insulator. *Science* **338**, 1193–1196 (2012).
12. Cheiwchanchamnangij, T. & Lambrecht, W. R. L. Quasiparticle band structure

calculation of monolayer, bilayer, and bulk $MoS_2$. *Phys. Rev. B* **85**, 205302 (2012).

13. Roldán, R. *et al.* Electronic properties of single-layer and multilayer transition metal dichalcogenides $MX_2$ (M = Mo, W and X = S, Se). *Ann. Phys.* **526**, 347–357 (2014).

14. Kormányos, A. *et al.* k·p theory for two-dimensional transition metal dichalcogenide semiconductors. *2D Mater.* **2**, 022001 (2015).

15. Neal, A. T., Liu, H., Gu, J. & Ye, P. D. Magneto-transport in $MoS_2$: Phase coherence, spin-orbit scattering, and the Hall factor. *ACS Nano* **7**, 7077–7082 (2013).

16. Zhou, Y., Yang, P., Zu, H., Gao, F. & Zu, X. Electronic structures and magnetic properties of $MoS_2$ nanostructures: atomic defects, nanoholes, nanodots and antidots. *Phys. Chem. Chem. Phys.* **15**, 10385–10394 (2013).

17. Zheng, H. *et al.* Tuning magnetism of monolayer $MoS_2$ by doping vacancy and applying strain. *Appl. Phys. Lett.* **104**, 132403 (2014).

18. Houssa, M., Iordanidou, K., Pourtois, G., Afanas'ev, V. V. & Stesmans, A. Point defects in $MoS_2$: Comparison between first-principles simulations and electron spin resonance experiments. *Appl. Surf. Sci.* **416**, 853–857 (2017).

19. Chiappe, D. *et al.* Controlled Sulfurization Process for the Synthesis of Large Area $MoS_2$ Films and $MoS_2/WS_2$ Heterostructures. *Adv. Mater. Interfaces* **3**, 1–10 (2016).

20. Marumoto, K., Kuroda, S. I., Takenobu, T. & Iwasa, Y. Spatial extent of wave functions of gate-induced hole carriers in pentacene field-effect devices as investigated by electron spin resonance. *Phys. Rev. Lett.* **97**, 256603 (2006).

21. Nagamori, T. & Marumoto, K. Direct observation of hole accumulation in polymer solar cells during device operation using light-induced electron spin resonance. *Adv. Mater*. **25**, 2362–2367 (2013).

22. Sato, G. *et al.* Direct observation of radical states and the correlation with performance degradation in organic light-emitting diodes during device operation. *Phys. Status Solidi A* **215**, 1700731 (2018).

23. Fujita, N. *et al.* Direct observation of electrically induced Pauli paramagnetism in single-layer graphene using ESR spectroscopy. *Sci. Rep.* **6**, 34966 (2016).

24. Pickard, C. J. & Mauri, F. First-Principles Theory of the EPR *g* Tensor in Solids: Defects in Quartz. *Phys. Rev. Lett.* **88**, 4 (2002).

25. Oshikawa, M. & Affleck, I. Electron spin resonance in $S$ = 1/2 antiferromagnetic chains. *Phys. Rev. B* **65**, 134410 (2002).

26. Heo, S., Hayakawa, R. & Wakayama, Y. Carrier transport properties of $MoS_2$ field-effect transistors produced by multi-step chemical vapor deposition method. *J. Appl. Phys.* **121**, 024301 (2017).

27. Jeong, Y. *et al.* Structural characterization and transistor properties of thickness-controllable $MoS_2$ thin films. *J. Mater. Sci.* **54**, 7758–7767 (2019).

28. Lee, J. *et al.* Ion gel-gated polymer thin-film transistors: Operating mechanism and characterization of gate dielectric capacitance, switching speed, and stability. *J. Phys. Chem. C* **113**, 8972–8981 (2009).

29. Pearce, A. J. & Burkard, G. Electron spin relaxation in a transition-metal dichalcogenide quantum dot. *2D Mater.* **4**, 025114 (2017).

30. Kormányos, A., Zólyomi, V., Drummond, N. D. & Burkard, G. Spin-orbit coupling, quantum dots, and qubits in monolayer transition metal dichalcogenides. *Phys. Rev. X* **4**, 011034 (2014).

31. Žutić, I., Fabian, J. & Sarma, S. Das. Spintronics : Fundamentals and applications. *Rev. Mod. Phys.* **76**, 323–410 (2004).

32. Mori, H. & Kawasaki, K. Theory of Dynamical Behaviors of Ferromagnetic Spins. *Prog. Theor. Phys.* **27**, 529–570 (1962).

33. Zhou, W. *et al.* Intrinsic structural defects in monolayer molybdenum disulfide. *Nano Lett.* **13**, 2615–2622 (2013).

34. Hwang, E. H., Adam, S. & Sarma, S. Das. Carrier transport in two-dimensional

graphene layers. *Phys. Rev. Lett.* **98**, 186806 (2007).

35. Lembke, D., Bertolazzi, S. & Kis, A. Single-layer $MoS_2$ electronics. *Acc. Chem. Res.* **48**, 100–110 (2015).

36. Ma, Y., Lehtinen, P. O., Foster, A. S. & Nieminen, R. M. Magnetic properties of vacancies in graphene and single-walled carbon nanotubes. *New J. Phys.* **6**, 68 (2004).

37. Panzer, M. J. & Frisbie, C. D. Exploiting ionic coupling in electronic devices: Electrolyte-gated organic field-effect transistors. *Adv. Mater.* **20**, 3176–3180 (2008).

38. Cho, J. H. *et al.* Printable ion-gel gate dielectrics for low-voltage polymer thin-film transistors on plastic. *Nat. Mater.* **7**, 900–906 (2008).

39. Pu, J. *et al.* Highly flexible $MoS_2$ thin-film transistors with ion gel dielectrics. *Nano Lett.* **12**, 4013–4017 (2012).

40. Giannozzi, P. *et al.* QUANTUM ESPRESSO: A modular and open-source software project for quantum simulations of materials. *J. Phys.: Condens. Matter* **21**, 395502 (2009).

41. Pickard, C. J. & Mauri, F. All-electron magnetic response with pseudopotentials: NMR chemical shifts. *Phys. Rev. B* **63**, 245101 (2001).

42. Blöchl, P. E. Projector augmented-wave method. *Phys. Rev. B* **50**, 17953–17979 (1994).

43. Pickard, C. J. & Mauri, F. Nonlocal Pseudopotentials and Magnetic Fields. *Phys. Rev. Lett.* **91**, 196401 (2003).

44. Marzari, N., Mostofi, A. A., Yates, J. R., Souza, I. & Vanderbilt, D. Maximally localized Wannier functions: Theory and applications. *Rev. Mod. Phys.* **84**, 1419–1475 (2012).

45. Pizzi, G. *et al.* Wannier90 as a community code: new features and applications. *J. Phys.: Condens. Matter* **32**, 165902 (2020).

46. Kokalj, A. Computer graphics and graphical user interfaces as tools in simulations of

matter at the atomic scale. *Comput. Mater. Sci.* **28**, 155–168 (2003).

Supplementary Information for

# $MoS_2$ thin-film transistors spin-states, of conduction electrons and vacancies, distinguished by operando electron spin resonance

Naho Tsunetomo, Shohei Iguchi, Małgorzata Wierzbowska, Akiko Ueda, Won Yousang, Heo Sinae, Jeong Yesul, Yutaka Wakayama, Kazuhiro Marumoto*

*To whom correspondence should be addressed; E-mail: marumoto@ims.tsukuba.ac.jp

**This PDF file includes:**

## A. Theory for the effective $g$-factor at finite temperature

Figure 4b shows the reduction of the effective $g$-factor with the increase of temperature. To understand the mechanism of the decrease, we consider a simple model for the monolayer $MoS_2$ including the spin relaxation. In the presence of the magnetic field, there can be two types of spin relaxation due to the spin-orbit interaction; the Elliot-Yaffet (EY) relaxation between the conduction and valence bands and the D'yakonov-Perel' (DP) relaxation inside the intra-band. Moreover, we consider the electron-phonon scattering which supports the spin relaxation.

The Hamiltonian of monolayer $MoS_2$ is written as $H_{\mathrm{MoS_2}} = H_{\mathrm{el}} + H_{\mathrm{zeeman}} + H_{\text{spin-relax}} + H_{\text{e-ph}}$. Here,

$$H_{\mathrm{el}} = \sum_{\boldsymbol{k},\sigma,\alpha=\mathrm{K,K'}} \varepsilon_{\boldsymbol{k},\alpha} c^{\dagger}_{\boldsymbol{k},\sigma,\alpha} c_{\boldsymbol{k},\sigma,\alpha} + \sum_{\boldsymbol{k},\sigma,\alpha=\mathrm{K,K'}} (\varepsilon_{\boldsymbol{k},\alpha} - E_{\mathrm{g}}) c^{\dagger}_{\boldsymbol{k},\sigma,\alpha,\mathrm{va}} c_{\boldsymbol{k},\sigma,\alpha,\mathrm{va}}, \quad (1)$$

$$H_{\mathrm{zeeman,K}} = g^{*} \mu_{\mathrm{B}} B_z S_z + \sum_{\boldsymbol{k},\sigma,\alpha} g_{\mathrm{v}} \mu_{\mathrm{B}} B_z \tau c^{\dagger}_{\boldsymbol{k},\sigma,\alpha} c_{\boldsymbol{k},\sigma,\alpha}. \quad (2)$$

$\varepsilon_{\boldsymbol{k},\alpha}$ is the energy of the electron with the momentum $\boldsymbol{k}$ at the valley $\alpha$ (= K or K') in the conduction band. $c^{\dagger}_{\boldsymbol{k},\sigma,\alpha}$ and $c_{\boldsymbol{k},\sigma,\alpha}$ are the creation and annihilation operators of electrons in the conduction band in the valley $\alpha$ with electron spin $\sigma\,(=\uparrow \text{ or } \downarrow)$, while $c^{\dagger}_{\boldsymbol{k},\sigma,\alpha,\mathrm{va}}$ and $c_{\boldsymbol{k},\sigma,\alpha,\mathrm{va}}$ are the creation and annihilation operators of electrons in the valence band in the valley $\alpha$, respectively. The spin operator in the $z$ direction is defined as $S_z = \frac{1}{2}\sum_{\boldsymbol{k},\alpha}\left(c^{\dagger}_{\boldsymbol{k},\uparrow,\alpha} c_{\boldsymbol{k},\uparrow,\alpha} - c^{\dagger}_{\boldsymbol{k},\downarrow,\alpha} c_{\boldsymbol{k},\downarrow,\alpha}\right)$. $g^{*}$ is the effective $g$-factor of spin at the zero temperature which includes the shift of $g$-factor due the intrinsic spin-orbit interaction of $MoS_2$. $g_{\mathrm{v}}$ denotes the valley $g$-factor. $\tau = 1\,(-1)$ for $\alpha = \mathrm{K}\,(\alpha = K')$. The Hamiltonian of the spin relaxation is expressed as

$$H_{\mathrm{spin-relax}} = \sum_{\boldsymbol{k},\sigma(\sigma\neq\sigma'),\alpha} L_{DP,\boldsymbol{k}} c^{\dagger}_{\boldsymbol{k},\sigma,\alpha} c_{\boldsymbol{k},\sigma',\alpha} + H.c. + \sum_{\boldsymbol{k},\sigma(\sigma\neq\sigma'),\alpha} L_{EY,\boldsymbol{k}} c^{\dagger}_{\boldsymbol{k},\sigma,\alpha} c_{\boldsymbol{k},\sigma',\alpha,\mathrm{va}} + H.c. \quad (3)$$

The first term indicates the DP type relaxation while the second term indicates the EY type relaxation[1]. The electron-phonon interaction in the conduction band is presented as

$$H_{\text{e-ph}} = \sum_{q,k,\sigma} M_{q,\mathrm{inter}} \left[c^{\dagger}_{k+q,\sigma,K} c_{k,\sigma,K'} \left(b_q + b^{\dagger}_q\right) + c^{\dagger}_{k+q,\sigma,K'} c_{k,\sigma,K} \left(b_q + b^{\dagger}_q\right)\right] +$$

$$\sum_{q,k,\sigma,\alpha} M_{q,\text{intra}} c^{\dagger}_{k+q,\sigma,\alpha} c_{k,\sigma,\alpha} \left(b_q + b^{\dagger}_q\right), \qquad (4)$$

where the first term denotes the inter-band phonon scattering between K and K' valleys while the second term denotes the intra-band scattering. The Hamiltonian for phonon is written as

$$H_{\text{ph}} = \sum_q \hbar\omega_q b^{\dagger}_q b_q. \qquad (5)$$

Several theoretical studies show that the optical phonon scatterings are the dominant factor among the electron-phonon scattering[2,3]. Thus, we only consider the optical phonon scattering for the electron-phonon scattering. We adopt the Einstein relation $\omega_q = \omega_0$. Accordingly, the electron-phonon coupling strength is written as

$$\sum_q M_{q,\text{inter(intra)}} = \frac{m^* D^2_{\text{inter(intra)}}}{2\varepsilon_m \hbar\omega_0}, \qquad (6)$$

where $D_{\text{inter(intra)}}$ is the deformation potential for intervalley (intravalley) scattering, and $\varepsilon_m$ is the ion mass density. To obtain the effective $g$-factor at finite temperature, we adopt the field-theory approach to the ESR. Using the Mori-Kawasaki formula, the ESR signal intensity is written as[1],

$$I(\omega) = \frac{B^2_{\perp}\omega}{2\mu_0} \chi_{\perp}(\omega) V, \qquad (7)$$

where $B_{\perp}$ is the magnetic induction of the electromagnetic radiation, $\mu_0$ is permeability of vacuum, and $V$ is the sample volume. Then the linear response of the spin susceptibility is

$$\chi_{\perp}(\omega) = -\text{Im}\, G^r_{S^+S^-}(\omega). \qquad (8)$$

where $G^r_{S^+S^-}(t,t') = -i\,\theta(t-t')\langle[S^+(t), S^-(t')]\rangle$. $\langle\cdots\rangle$ denotes the quantum and statistical average over states of the whole system. Here, $S^+(t) = \sum_\alpha S_{x,\alpha} \pm iS_{y,\alpha} = \sum_{\boldsymbol{k},\alpha} c^{\dagger}_{\boldsymbol{k},\uparrow,\alpha} c_{\boldsymbol{k},\downarrow,\alpha}$ and $S^-(t) = \sum_\alpha S_{x,\alpha} \pm iS_{y,\alpha} = \sum_{\boldsymbol{k},\alpha} c^{\dagger}_{\boldsymbol{k},\downarrow,\alpha} c_{\boldsymbol{k},\uparrow,\alpha}$. Using the equation of motion, we obtain[1,4]

$$G^r_{S^+S^-}(\omega) = \frac{2\hbar\langle S_z\rangle}{\hbar\omega - g\mu_{\text{B}}B_z - \Sigma(\omega)}. \qquad (9)$$

Here, the self-energy is written as

$$\Sigma(\omega) = \frac{-\langle[A(0), S^{-}(0)]\rangle + G^{r}_{AA^{\dagger}}(t,t')}{2\langle S_z\rangle}, \qquad (10)$$

where $A = [H_{\text{spin-relax}}, S^{+}]$, and $G^{r}_{AA^{\dagger}}(t,t') = -i\,\theta(t-t')\langle[A(t), A^{\dagger}(t')]\rangle$.

To calculate $G^{r}_{AA^{\dagger}}(\omega)$, we only take into account the second-order in the spin relaxation. The electron-phonon interaction is included in the second order in the self-energy of the single-particle Green's function $G^{r}_{\boldsymbol{k}\sigma\alpha,\boldsymbol{k}\sigma\alpha}(t,t') = -i\,\theta(t-t')\langle[c_{\boldsymbol{k},\sigma,\alpha}(t), c^{\dagger}_{\boldsymbol{k},\sigma,\alpha}(t')]\rangle$.

The peak energy of the ESR resonance $\hbar\omega_{\text{peak}}$ is derived by

$$\hbar\omega_{\text{peak}} + g^{*}\mu_B B_z - \text{Re}\Sigma(\omega) = 0. \qquad (11)$$

Hence, the shift of the $g$-factor due to the spin relaxation and the e-ph interaction is

$$\Delta g = (\hbar\omega_{\text{peak}} + g^{*}\mu_{\text{B}} B_z)/\mu_B B_z. \quad (12)$$

To fit the theoretical result with the experimental result of Figure 4b, we assume the parameters in the Table S1 below. $g^{*}$, $|L_{DP,k}|^{2}$, and $|L_{EY,k}|^{2}$ are fitted with experimental result and the other parameters are taken from references presented. We also assume that the gate with the ion gel shifts the Fermi energy, $\varphi_n$, above the conduction band, $E_{\text{c}}$ so that $\varphi_n - E_{\text{c}} = 5$ meV. The magnetic field is fixed to $B_z = -0.32$ T. In Figure S1, we plot $g(T) = g^{*} + \Delta g$ as a function of the temperature $T$. $g(T)$ decreases due to the spin-relaxation and well match with the experimental result.

**Table S1 | Parameters for fitting the experimental result in Figure 4b.**

| Parameter | Symbol | Value |
|---|---|---|
| *g*-factor for spin in the absence of spin relaxation | $g^*$ | 2.077 |
| DP spin relaxation energy | $\left\|L_{DP,k}\right\|^2$ | $5 \times 10^{-7}$ (meV)$^2$ |
| EY spin relaxation energy | $\left\|L_{EY,k}\right\|^2$ | $1.6 \times 10^{-4}$ (meV)$^2$ |
| *g*-factor of valleys | $g_v$ | 3.57 (ref. 5) |
| Phonon energy | $\hbar\omega_0$ | 50 meV (ref. 2) |
| Ion mass density | $\varepsilon_m$ | $3.1 \times 10^{-7}$ g/cm$^2$ (ref. 3) |
| Deformation potential (inter) | $D_{\text{inter}}$ | $2.0 \times 10^8$ eV/m (ref. 3) |
| Deformation potential (intra) | $D_{\text{intra}}$ | $5.8 \times 10^8$ eV/m (ref. 3) |
| effective mass | $m^*$ | $0.48 \times m_0$ ($\sim 9.1 \times 10^{-31}$kg) (ref. 2) |
| Electron-phonon coupling (inter band) | $\sum_q M_{q,\text{inter}} = \frac{m^* D_{\text{inter}}^2}{2\varepsilon_m \hbar\omega_0}$ | ~0.56 meV |
| Electron-phonon coupling (intra band) | $\sum_q M_{q,\text{intra}} = \frac{m^* D_{\text{intra}}^2}{2\varepsilon_m \hbar\omega_0}$ | ~4.7 meV |
| Energy bandgap | $E_g$ | 1.85 eV (ref. 5) |

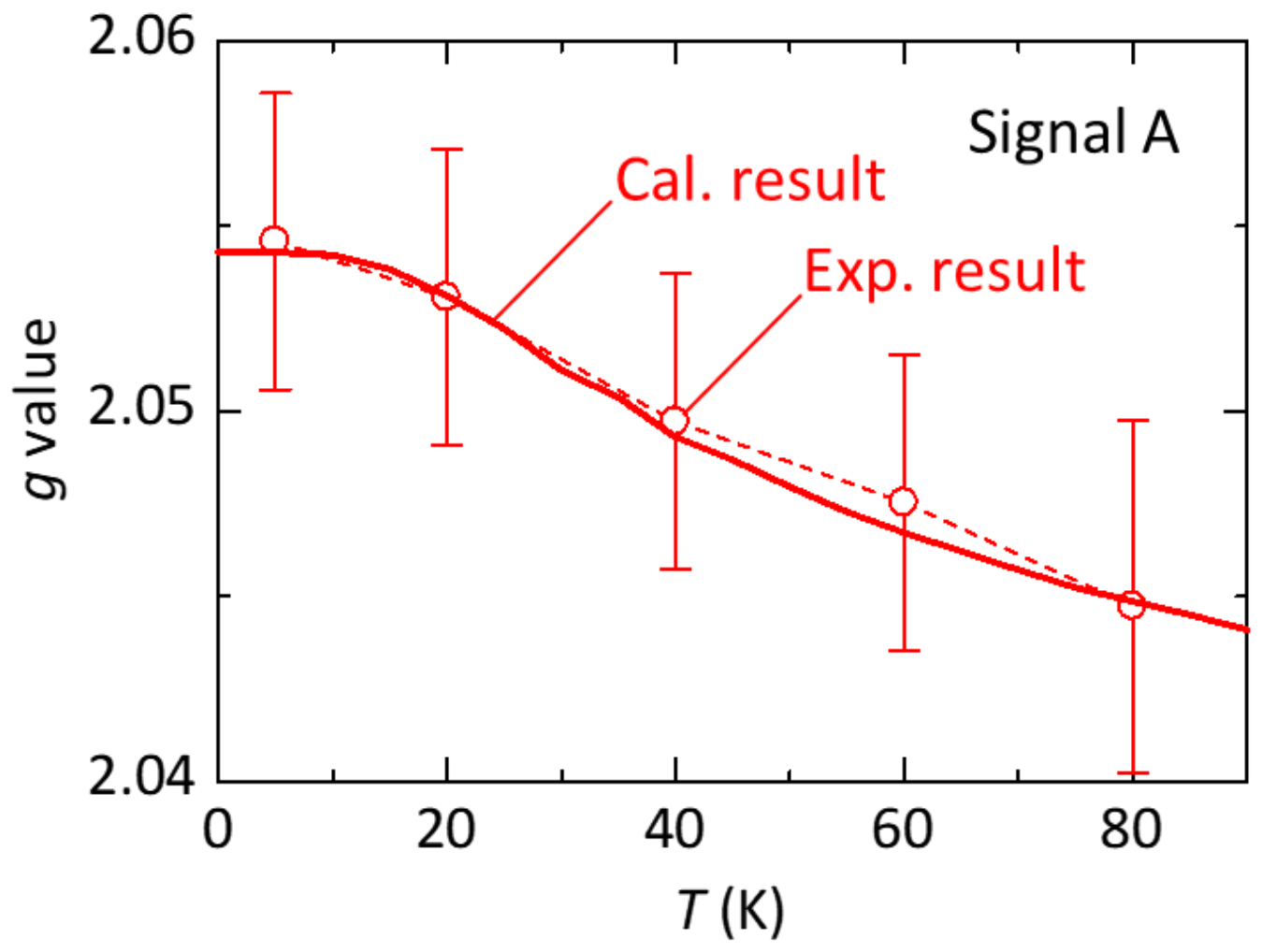

**Figure S1 | *g*-factor as a function of temperature.** Solid line indicates the numerical result calculated by the Mori-Kawasaki formula. The circles and error bars show the experimental result in Figure 4b. The parameters in the Table S1 are used to fit the experimental data.

**B. Examples of calculated spin-density distribution around vacancies in the monolayer $MoS_2$**

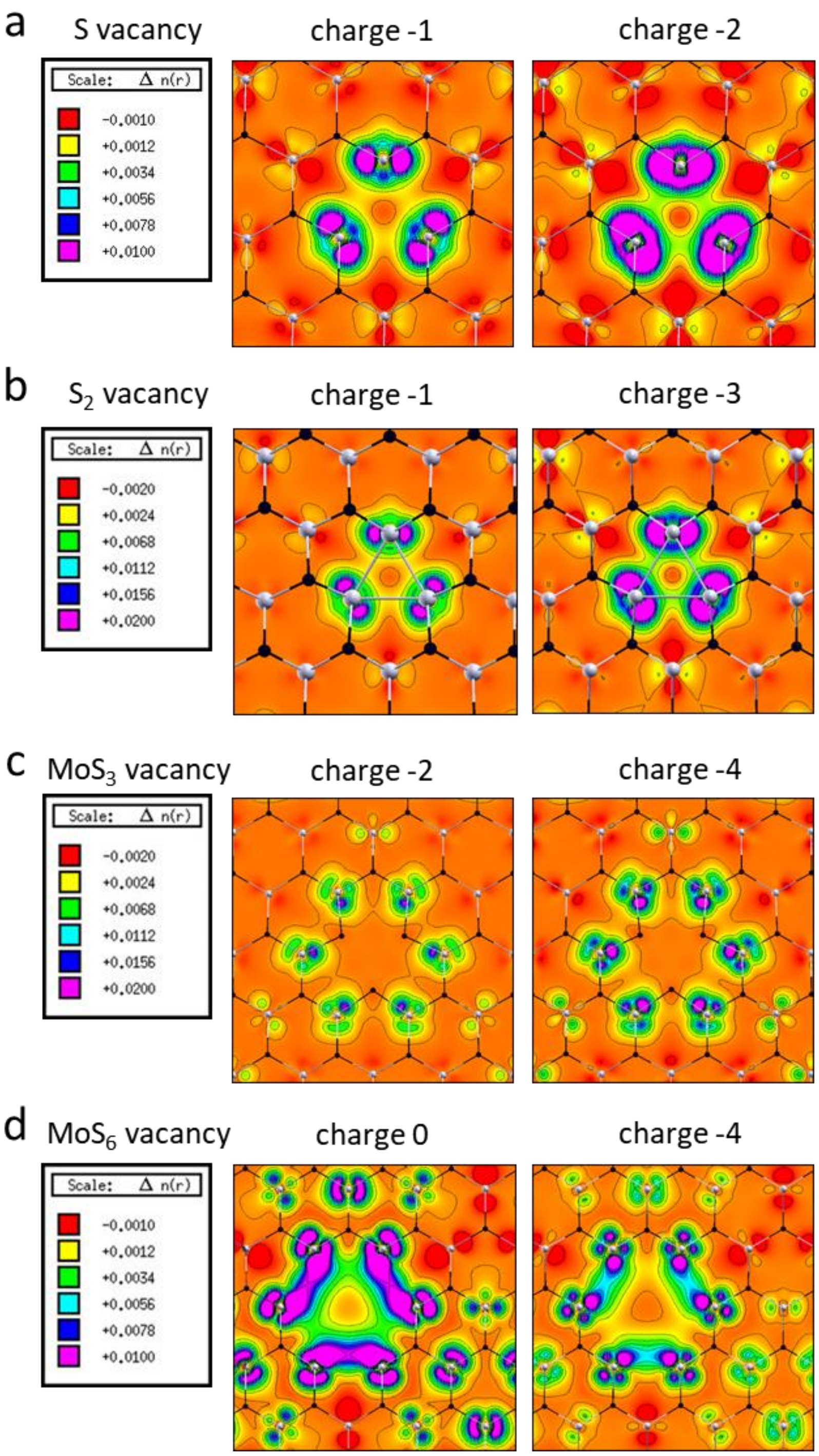

**Figure S2 | Maps of spin-density distribution for monolayer (mL) $MoS_2$ with various vacancies and charge doping**. **a**, mL $MoS_2$ with S vacancy. **b**, mL $MoS_2$ with $S_2$ vacancy. **c**, mL $MoS_2$ with $MoS_3$ vacancy. **d**, mL $MoS_2$ with $MoS_6$ vacancy.

**C. Definition of 3 types of S and $MoS_3$ vacancy configurations in the bilayer $MoS_2$**

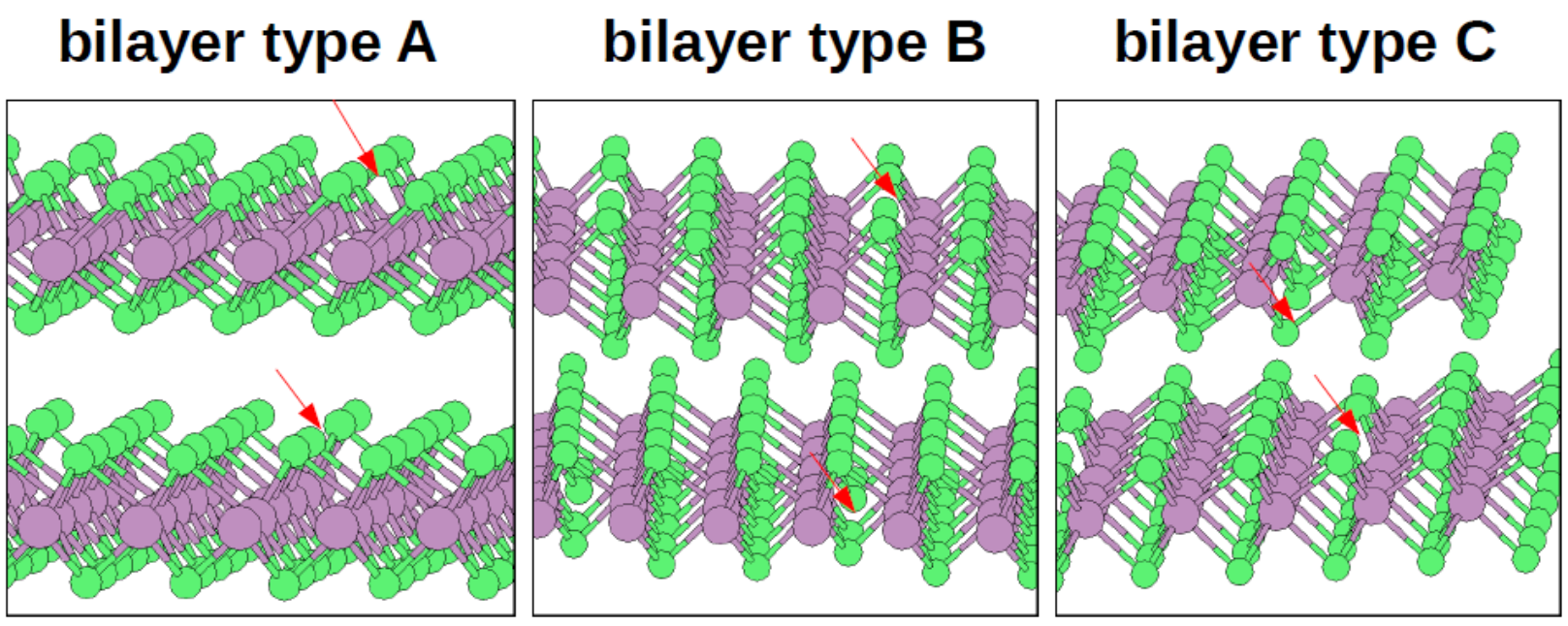

**Figure S3 | Schematics of 3 types of S and $MoS_3$ vacancy configurations in the bilayer $MoS_2$.** When the S vacancy enters into the bilayer $MoS_2$, three types are conceivable depending on how the vacancy enters. Type A: one vacancy is inside and the other is outside. Type B: both vacancies are outside of two $MoS_2$ monolayers. Type C: both vacancies are inside of two monolayer sheets of $MoS_2$.

## D. Calculated magnetization of $MoS_2$ thin films

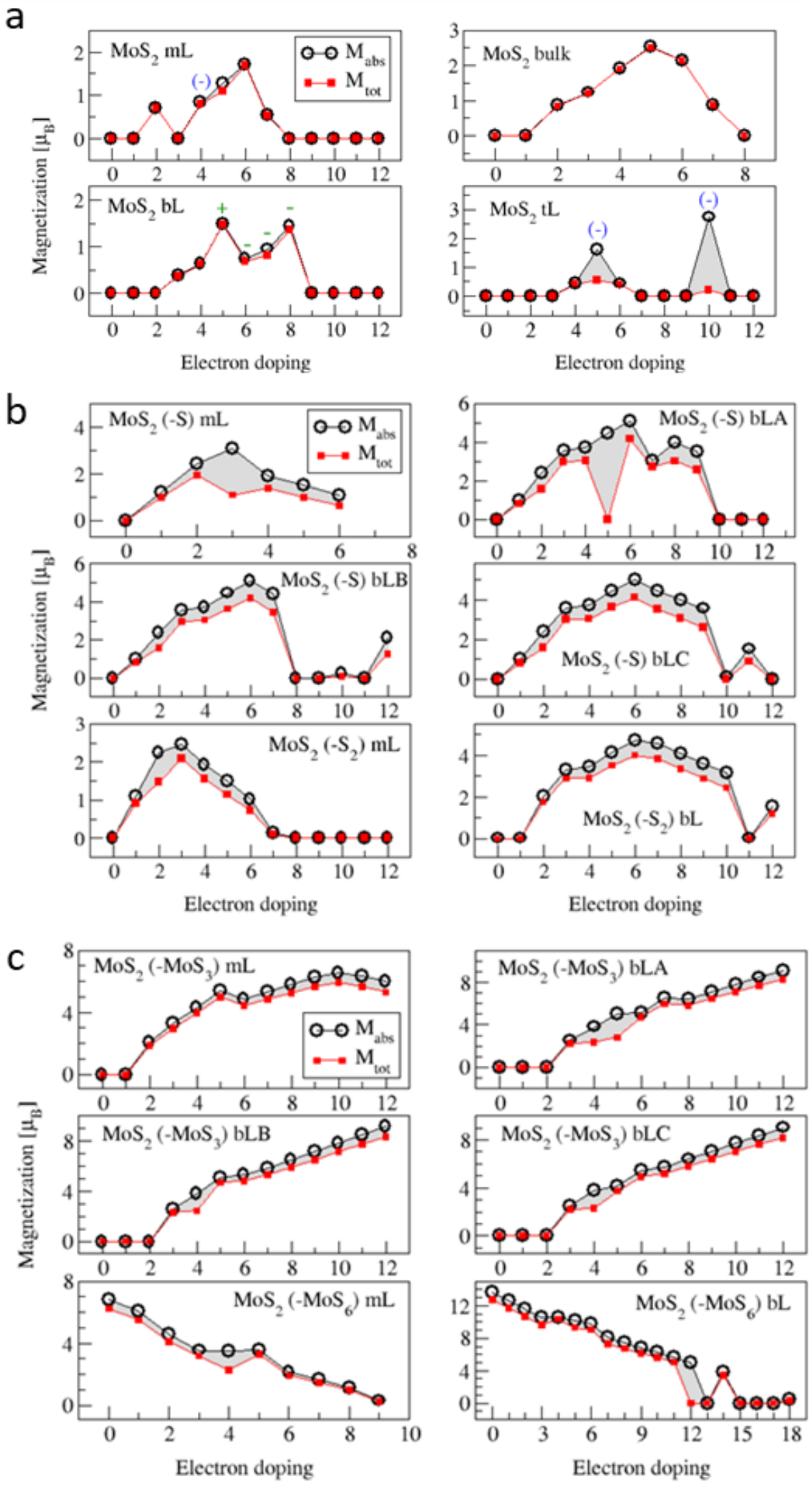

**Figure S4 | Electron-doping dependence of the magnetization of $MoS_2$ thin films.** $M_{abs}$ is the absolute magnetization defined as $\int |n_{up}(r) - n_{down}(r)|\, dr$, where $n_{up}(r)$ and $n_{down}(r)$ are the spin densities. $M_{tot}$ is the total magnetization defined as $\int n_{up}(r) - n_{down}(r)\, dr$. The terms of mL, bL and tL mean a monolayer, bilayer and trilayer, respectively. We calculated also the bulk material. **a**, Magnetization of mL, bL, tL and bulk $MoS_2$ with no vacancy. **b**, Magnetization of $MoS_2$ with S or $S_2$ vacancy. **c**, Magnetization of $MoS_2$ with $MoS_3$ or $MoS_6$ vacancy.

## E. Calculated spin-density distribution of $MoS_2$ with various vacancies and doping

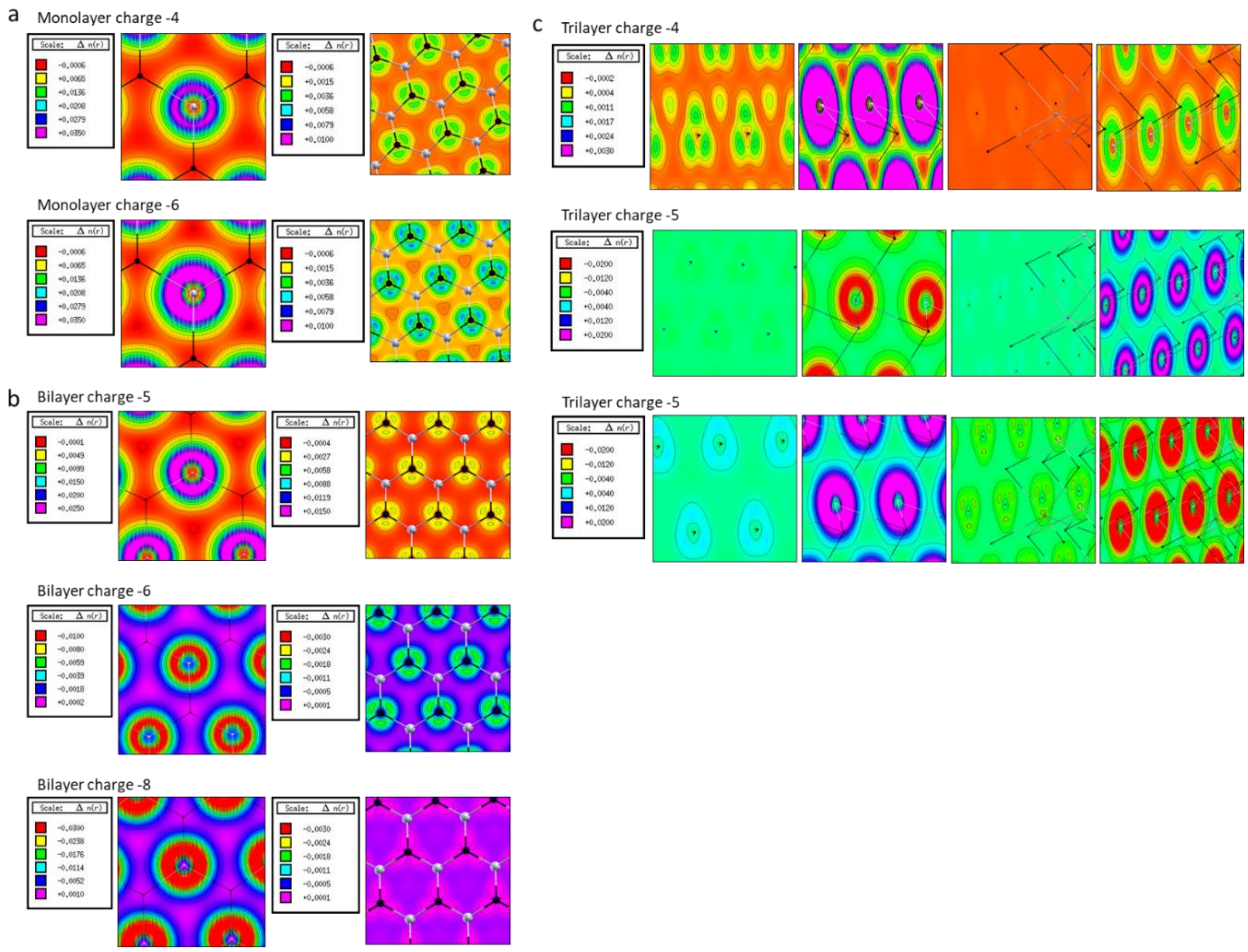

**Figure S5 | Maps of spin-density distribution of $MoS_2$ with no vacancy with various electron doping.** White and black balls represent Mo and S atoms, respectively. **a**,**b**, Maps on the left-hand and right-hand side show the planes which cut through the Mo and S atoms, respectively. **c**, Maps for the trilayer $MoS_2$ spin-density distributions on different planes which cut through the Mo or S atoms.

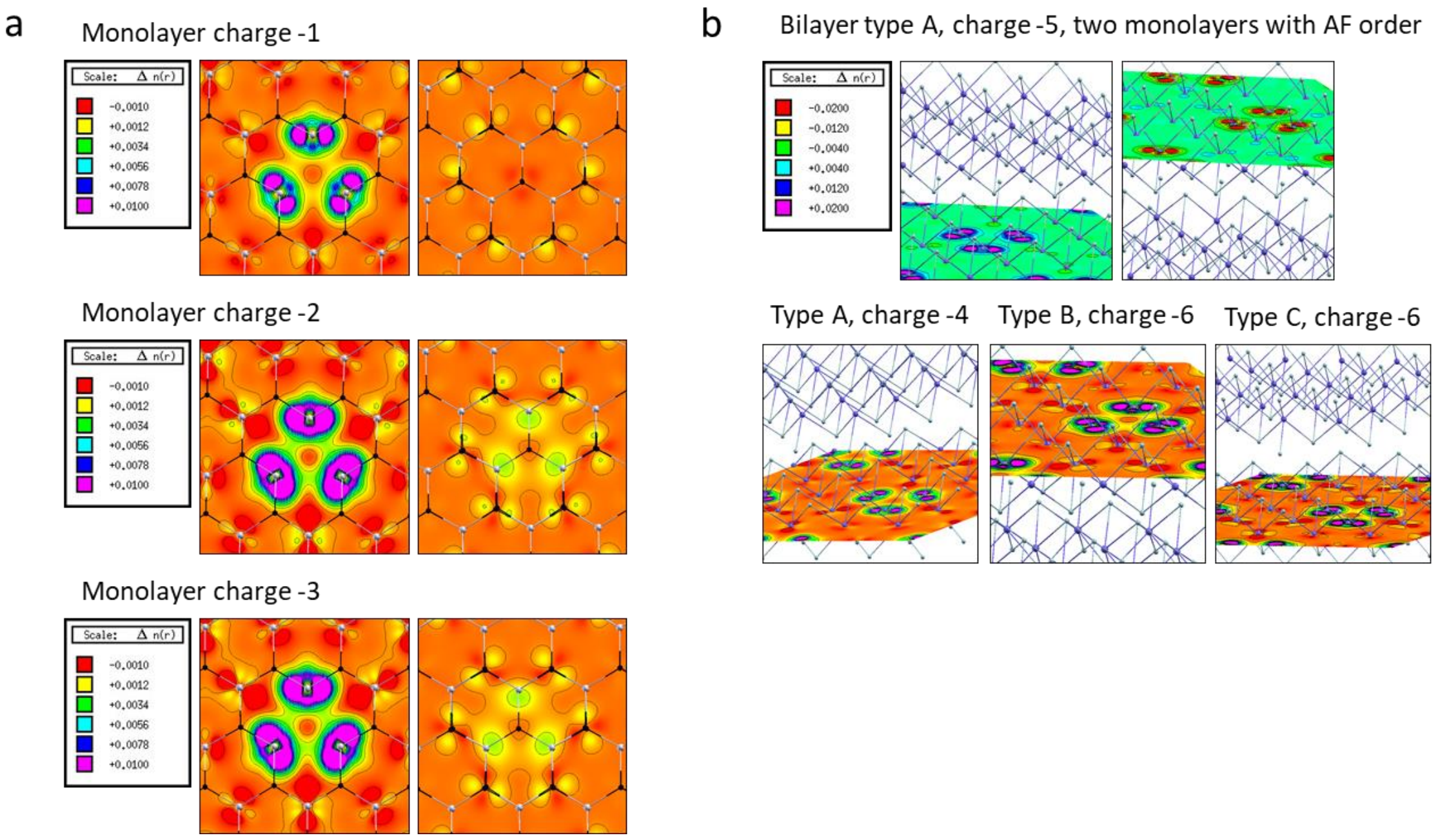

**Figure S6 | Maps of spin-density distribution of $MoS_2$ with S vacancy for monolayer and bilayer with various electron doping.** White and black balls represent Mo and S atoms, respectively. Maps on the left-hand and right-hand side show the planes which cut through the Mo and S atoms, respectively. **a**, Monolayer $MoS_2$ with various charges. **b**, Bilayer $MoS_2$ with different vacancy configurations (defined in Figure S3) and various charges. In (b), the antiferromagnetic (AF) order is represented.

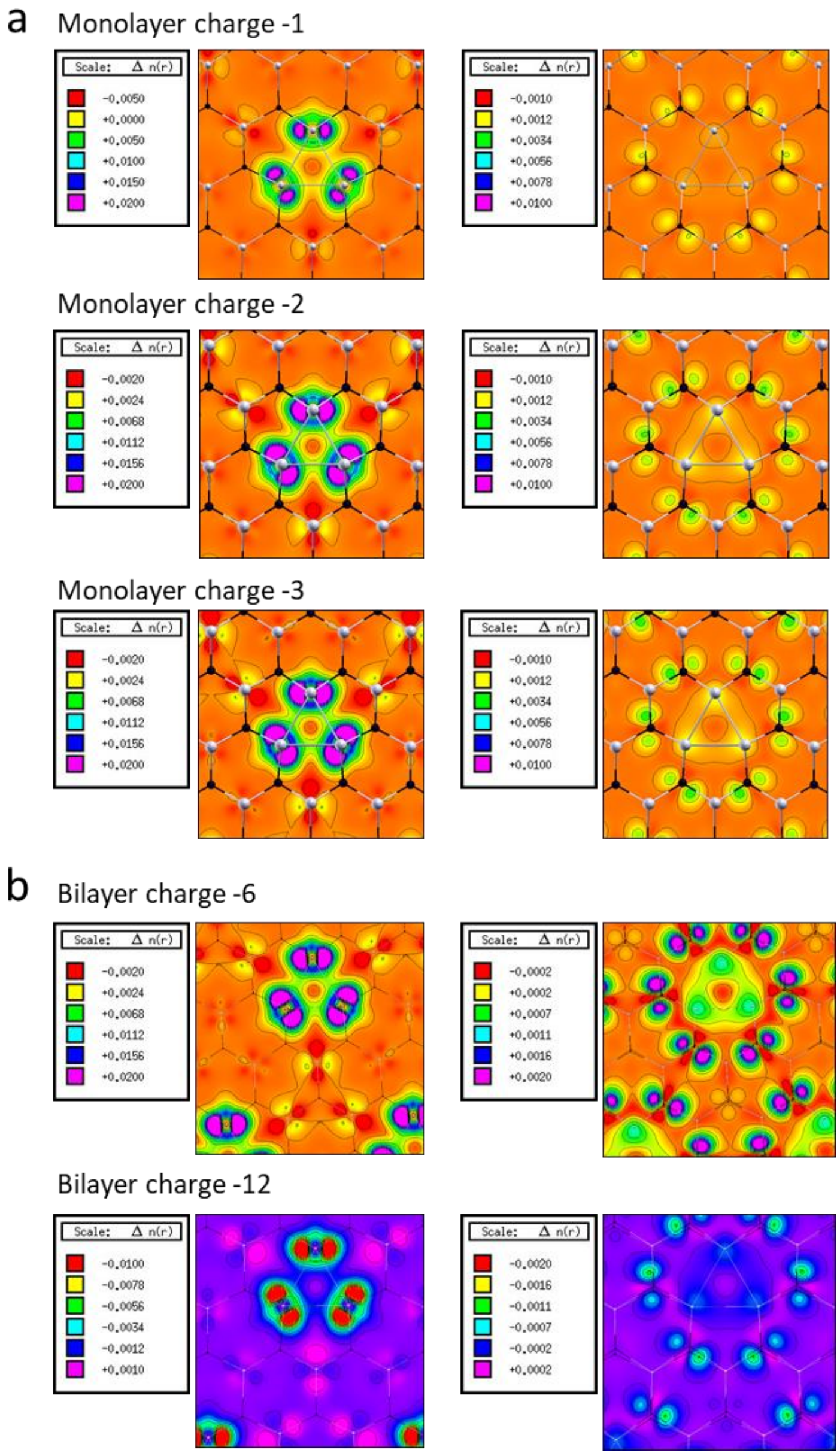

**Figure S7 | Maps of spin-density distribution of $MoS_2$ with $S_2$ vacancy for monolayer and bilayer with various electron doping** Maps on the left-hand and right-hand side show the planes which cut through the Mo and S atoms, respectively. **a**, Monolayer $MoS_2$ with $S_2$ vacancy with various charges. **b**, Bilayer $MoS_2$ with $S_2$ vacancy with various charges.

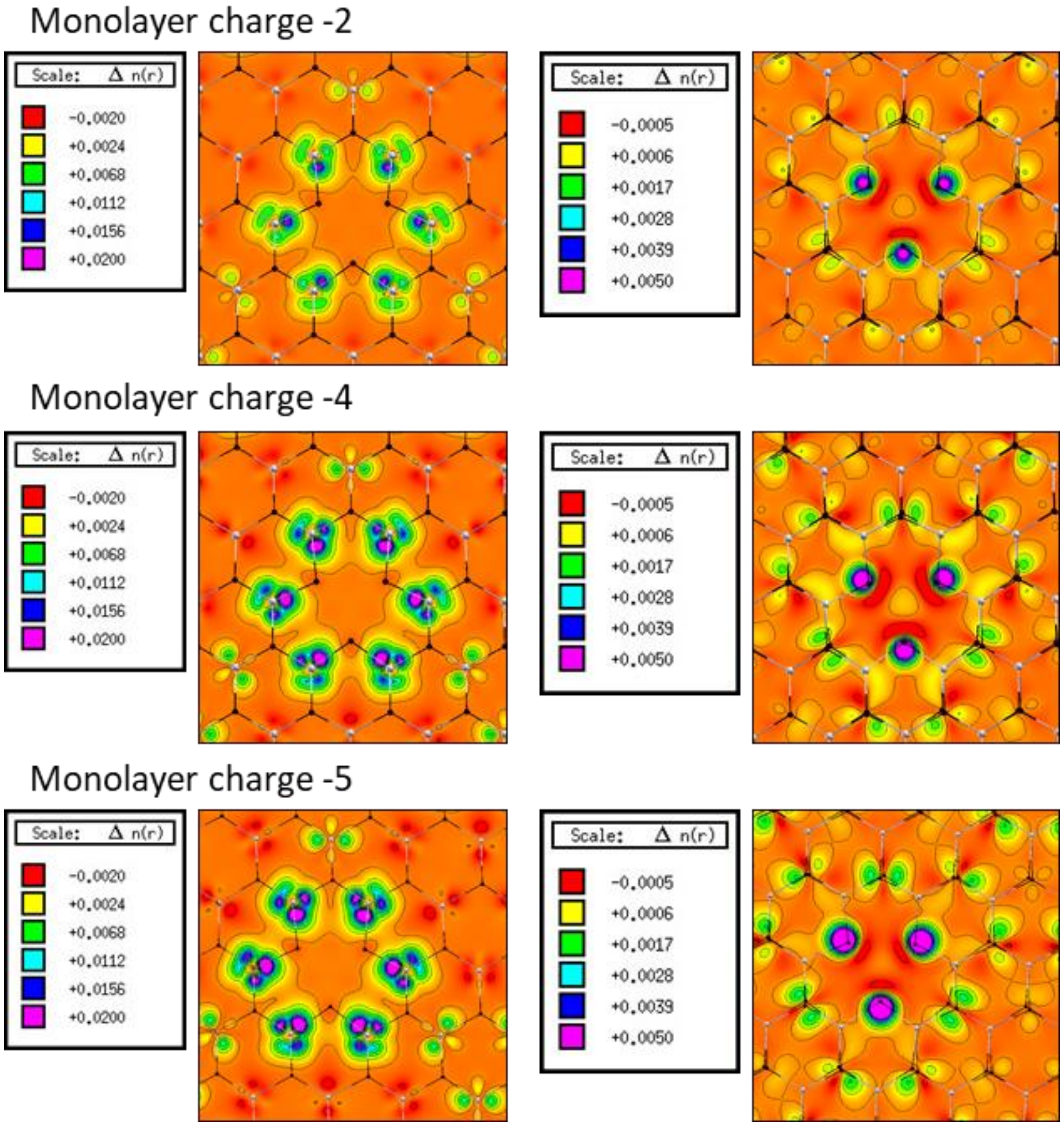

**Figure S8 | Maps of spin-density distribution of $MoS_2$ with $MoS_3$ vacancy for monolayer with various electron doping.** Maps on the left-hand and right-hand side show the planes which cut through the Mo and S atoms, respectively.

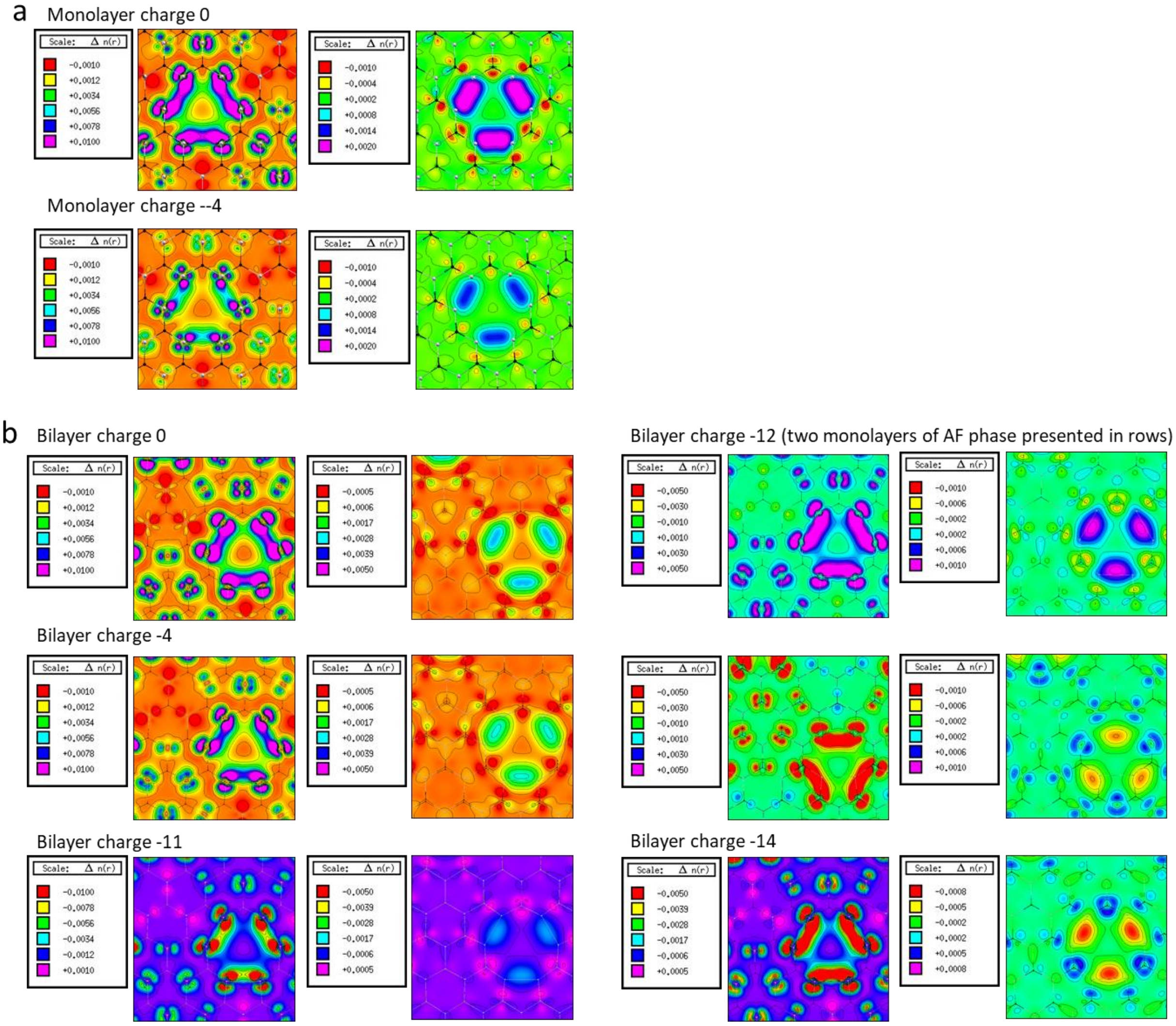

**Figure S9 | Maps of spin-density distribution of $MoS_2$ with $MoS_6$ vacancy for monolayer and bilayer with various electron doping.** Maps on the left-hand and right-hand side show the planes which cut through the Mo and S atoms, respectively. **a**, Monolayer $MoS_2$ with $MoS_6$ vacancy with various charges. **b**, Bilayer $MoS_2$ with $MoS_6$ vacancy with various charges. Unlike the other vacancies (S, $S_2$ and $MoS_3$ vacancies), the case of $MoS_6$ vacancy is magnetic in the charge neutral state. In (b), the antiferromagnetic (AF) order is represented for the case of charge −12.

### F. Calculated *g*-factors with various electron doping

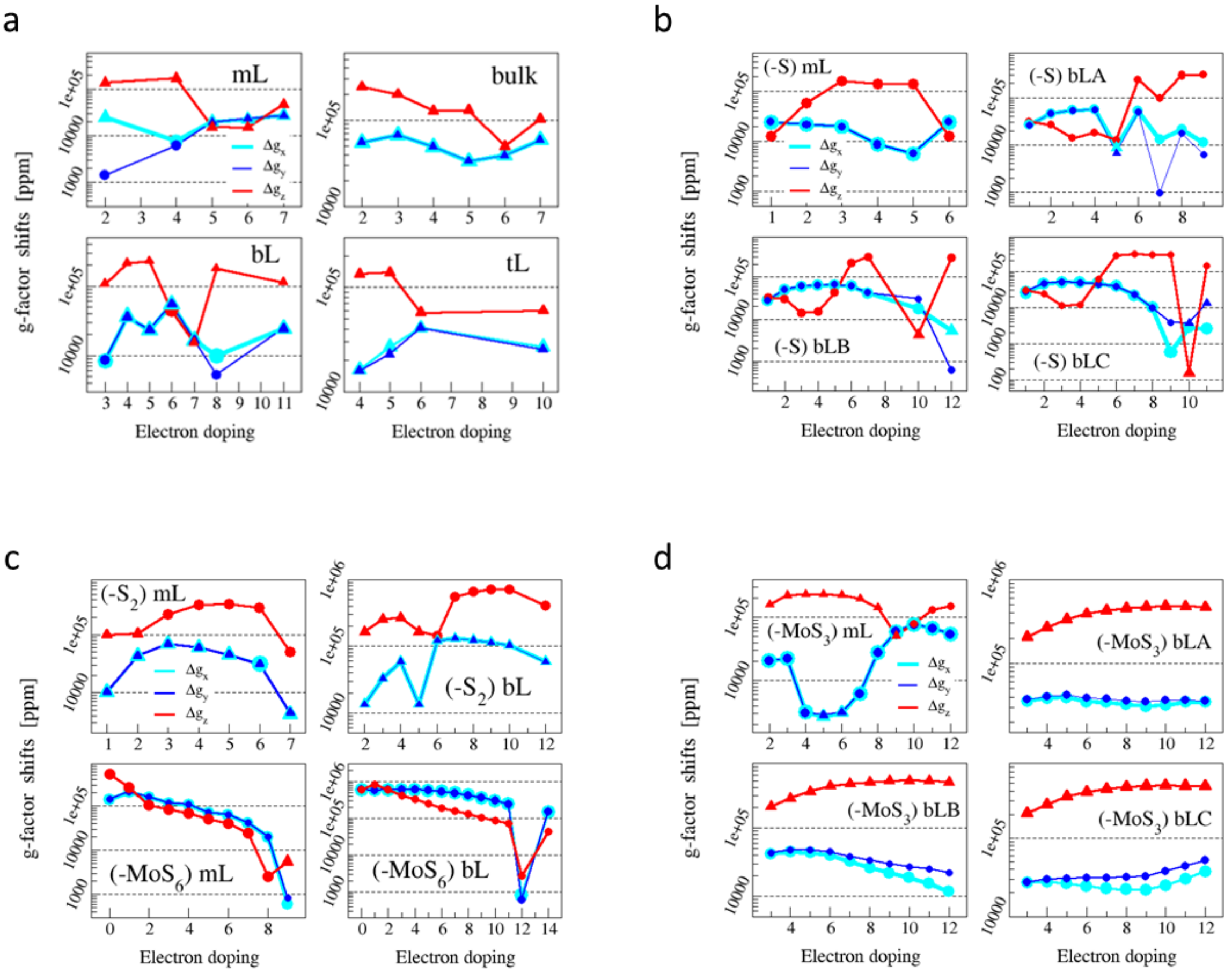

**Figure S10 | Shifts of *g*-factor from the free-electron and their dependence on the electron doping of the supercell.** The lines of light blue, blue and red show the calculated results for the *x*-, *y*- and *z*-axis direction, respectively, where *x*- and *y*-axes are paralleled to the $MoS_2$ plane and the *z*-axis is perpendicular to the $MoS_2$ plane. The shift amounts from the free-electron's value ($g$ = 2.0023) are shown. Circle and triangle data indicate positive and negative $g$ shift with absolute value, respectively. **a**, $MoS_2$ with no vacancy. **b**, $MoS_2$ with S vacancy. **c**, $MoS_2$ with $S_2$ or $MoS_6$ vacancy. **d**, $MoS_2$ with $MoS_3$ vacancy.

**References**

1. Boross, P., Dóra, B., Kiss, A. & Simon, F. A unified theory of spin-relaxation due to spin-orbit coupling in metals and semiconductors, *Sci. Rep.* **3**, 3233 (2013).

2. Kaasbjerg, K., Thygesen, K. S. & Jacobsen, K. W. Phonon-limited mobility in n-type single-layer $MoS_2$ from first principles, *Phys. Rev. B* **85**, 115317 (2012).

3. Li, X. *et al.*, Intrinsic electrical transport properties of monolayer silicene and $MoS_2$ from first principles, *Phys. Rev. B* **87**, 115418 (2013).

4. Oshikawa, M. & Affleck, I. Electron spin resonance in $S = \frac{1}{2}$ antiferromagnetic chains, *Phys. Rev. B* **65**, 134410 (2002).

5. Kormányos, A., Zólyomi V., Drummond, N. D. & Burkard, G. Spin-Orbit Coupling, Quantum Dots, and Qubits in Monolayer Transition Metal Dichalcogenides, *Phys. Rev. X* **4**, 011034 (2014).